\documentclass[useAMS,usenatbib]{mn2e} 

\usepackage{amssymb}
\usepackage{amsmath}
\usepackage{graphicx}
\usepackage{textcomp}
\usepackage{natbib}
\usepackage{Times}

\def\be{\begin{equation}}
\def\ee{\end{equation}}
\def\bea{\begin{eqnarray}}
\def\eea{\end{eqnarray}}

\title{Probing deviations from General Relativity with the Euclid spectroscopic survey}
\author[Elisabetta Majerotto et al.]{
  \parbox{\textwidth}{
    	E. Majerotto$^{1,2,3}$\thanks{E-mail: elisabetta.majerotto@uam.es},
    	L. Guzzo$^{1}$,
    	L. Samushia$^{4,5}$,
    	W. J. Percival$^{4}$,
    	Y. Wang$^{6}$,
	S. de la Torre$^{1,7}$,
	B. Garilli$^{8,9}$,
	P. Franzetti$^{8}$,
	E. Rossetti$^{10}$,
	A. Cimatti$^{10}$,
	C. Carbone$^{10,11,12}$,
	N. Roche$^{11}$,
	G. Zamorani$^{11}$
 }
 \vspace{0.4cm}\\
\parbox{\textwidth}{
$^{1}$  INAF-Osservatorio Astronomico di Brera, Via Emilio Bianchi, 46, I-23807,
  Merate (LC), Italy\\
  $^{2}$ Instituto de F{\'i}sica Te{\'o}rica (UAM/CSIC), Universidad Aut{\'o}noma de Madrid, Cantoblanco, 28049 Madrid, Spain\\
$^{3}$ Departamento de F{\'i}sica Te{\'o}rica (UAM), Universidad Aut{\'o}noma de Madrid, Cantoblanco, 28049 Madrid, Spain\\
  $^{4}$ Institute of Cosmology and Gravitation, University of
  Portsmouth, Dennis Sciama building, Portsmouth, P01 3FX, UK\\
  $^{5}$ National Abastumani Astrophysical Observatory, Ilia State University, 2A
  Kazbegi Ave, GE-0160 Tbilisi, Georgia\\
  $^6$ Department of Physics \& Astronomy University of Oklahoma,
  Norman, OK 73019, USA\\
  $^7$ Institute for Astronomy, University of Edinburgh, Royal Observatory,
Blackford Hill, EH9 3HJ Edinburgh, UK
  \\
  $^8$ IASF-Milano, INAF, Via Bassini, 15, I-20133, Milano, Italy\\
  $^9$ Laboratoire d'Astrophysique de Marseille, UMR 6110 CNRS-Universit{\'e} de Provence, 38 rue Frederic Joliot-Curie, F-13388\\ Marseille Cedex 13, France\\
  $^{10}$ Dipartimento di Astronomia, Alma Mater Studiorum-Universit{\`a} di Bologna, Via
Ranzani, 1, I-40127 Bologna, Italy\\
$^{11}$ INAF-Osservatorio Astronomico di Bologna,
Via Ranzani, 1, 40127 Bologna, Italy\\
  $^{12}$ INFN, Sezione di Bologna, Viale Berti Pichat, 6/2, I-40127 Bologna, Italy
  }}

\begin{document}

\date{\today}

\maketitle

\pagerange{\pageref{firstpage}--\pageref{lastpage}} \pubyear{2012}
\label{firstpage}
\begin{abstract}
We discuss the ability of the planned Euclid mission to detect
deviations from General Relativity using its extensive redshift survey
of more than 50 Million galaxies. Constraints on the gravity theory
are placed measuring the growth rate of structure within 14 redshift
bins between $z=0.7$ and $z=2$. The growth rate is measured from
redshift-space distortions, i.e. the anisotropy of the clustering
pattern induced by coherent peculiar motions. This is performed in the
overall context of the Euclid spectroscopic survey, which will
simultaneously measure the expansion history of the universe, using
the power spectrum and its baryonic features as a standard ruler,
accounting for the relative degeneracies of expansion and growth
parameters.  The resulting expected errors on the growth rate in the
different redshift bins, expressed through the quantity $f\sigma_8$,
range between $1.3\%$ and $4.4\%$.  We discuss the optimisation of the
survey configuration and investigate the important dependence on the
growth parameterisation and the assumed cosmological model.  We show
how a specific parameterisation could actually drive the design
towards artificially restricted regions of the parameter 
space.  Finally, in the framework of the popular  ``$\gamma$
parameterisation", we show that the Euclid spectroscopic survey alone
will already be able to provide {\it substantial evidence} (in
Bayesian terms) if the growth index differs from the GR value
$\gamma=0.55$ 
by at least $\sim 0.13$. 
This will combine with the comparable inference power provided by the
Euclid weak lensing survey, resulting in Euclid's unique ability to
provide a decisive test of modified gravity. 
\end{abstract}

\begin{keywords}
cosmology --  dark energy -- large-scale structure of universe.
\end{keywords}

\section{Introduction}
The current standard cosmological model, concordantly supported by
virtually all available observations, tells us that we live in a
low-density, expanding universe with a spatially flat geometry, that
appears to have recently entered a phase of accelerated expansion.
The data require an extra mass-energy
contribution in the form of a fluid with equation of state $w \sim -1$.
This corresponds to adding in the
equations of General Relativity (GR) a {\it cosmological constant} $\Lambda$, i.e. the term originally
introduced by Einstein to obtain a static solution, thus building the standard Lambda Cold Dark Matter model (LCDM). A cosmological constant,
 however, has a few disturbing
features. The first feature is the fine tuning necessary to obtain its measured density value, which can be interpreted as the energy of the vacuum, that is extremely small compared to the corresponding scales of particle physics; the second issue is the so-called coincidence problem: why, despite very different time evolutions, do dust-like matter and cosmological constant show comparable densities today? Among possible solutions, scenarios with evolving `dark
energy' density from a cosmological scalar field, have been
proposed (see \citealt{Fr} for a review).  Alternatively,
however, observations could simply indicate that it is the theory
of gravity that needs to be revised (see~\citealt{Co} for a review).

These two radically different explanations cannot be distinguished by
measuring only the expansion history of the universe represented by the Hubble function $H(z)$.  A way to break this
degeneracy is to look at the linear growth rate of density
perturbations.  This can be expressed as $f = d $ln$ G/d $ln$ a$,
where $G(t)$ is the time-dependent part of the solution of the linear
growth equation \citep{He} and $a$ is the cosmic scale
factor. Models with the same expansion history $H(z)$, but based on a
different gravity theory predict a different growth rate $f(z)$ \citep{Maartens:2006qf, Linder:2005in, Polarski:2006ut} (although see also \citealt{Kunz:2006ca} for possible issues). 

Measurements of galaxy clustering from large redshift surveys,
quantified through the galaxy-galaxy correlation function (or its Fourier transform, the power spectrum $P(k)$), contain direct information on
both $H(z)$ and $f(z)$.  

Baryonic Acoustic Oscillations (BAO) within
the last-scattering surface give rise to a characteristic feature in
the galaxy distribution at comoving separations of $\sim 150$
Mpc (BAO have now been conclusively seen in the clustering of
galaxies, e.g. \citealt{Cole:2005sx, Ei} and more recently  \citealt{Kazin:2009cj}).
This corresponds to the characteristic scale fixed by the comoving
sound horizon at the drag epoch (shortly after 
recombination) and is accurately measured by Cosmic
Microwave Background (CMB) anisotropies (see recent measurements by
\citealt{Ko}). Compared to the observed 
galaxy BAO peak position at different redshifts, this yields a
measurement of $H(z)$ in the radial direction and of the comoving
angular diameter distance $D_A(z)$ in the transverse direction. 

Galaxy clustering as measured in redshift space also contains the
imprint of the linear growth rate of structure, in the form of a
measurable anisotropy.  Such {\it redshift-space distortion} (RSD) is
due to coherent flows of matter from low to high densities, 
linked to the growth of structure. When redshifts are used to
measure galaxy distances, the contribution from peculiar velocities
produces a distortion of the clustering pattern, which at linear
scales is proportional to $f(z)$ \citep{Kaiser:1987qv, Ha}. This can
be measured by modelling the anisotropy of redshift-space two-point
statistics, defining a proper correlation function $\xi(r_p , \pi)$ (or
power spectrum $P(k_\parallel,k_\perp)$),
where $r_p$ and $k_\parallel$ ($\pi$ and $k_\perp$) are the components parallel (perpendicular) to the line of sight. 
The anisotropy is caused by an additive
term proportional to the variance of the velocity field, which can be
parameterised by $f(z)\sigma_8(z)$ (where $\sigma_8$ is the rms
amplitude of galaxy clustering) and provides an excellent
discriminator of cosmological models, particularly if $\sigma_8$ is
normalised, for example using the CMB \citep{Song:2008qt}.   

RSD were classically seen as a method to measure $\Omega_0$ (see
references in \citet{Ha} or more recently \citet{Hw}, \citet{Ro},
\citet{daAngela:2006mf}, \citet{CGA}, \citet{Drinkwater:2009sd}). In
the dark energy context, they were initially seen as simply an effect to be
corrected for to extract the full BAO
information \citep{Se}. Still in the context of GR,
\citet{Amendola:2004be} showed that the information contained in the GR
growth function $f(z)$  could improve errors on $w(z)$
parameters by $\sim 30\%$ (see also \citealt{Sapone:2007jn}). As pointed out by \citet{Guzzo:2008ac}, however, if no assumption is made on the gravity theory, RSD in fact provide us with a powerful test to test the dark energy vs modified gravity alternative by tracing the growth rate back in time. This was particularly interesting in the context of planned dark energy surveys, and stimulated new interest in this technique \citep{Wang:2008, Li2, Ne, Ac, Song:2008qt, WSP, PW}. Following
\citet{Guzzo:2008ac},  RSD were suggested as a primary probe in the
SPACE satellite proposal (the forerunner of what is now the Euclid
spectroscopic probe, described in more detail below) presented in response to the 2007 ESA  
Cosmic Vision Call \citep{Ci}.

Currently ongoing spectroscopic surveys such as WiggleZ
\citep{Blake:2011rj}, BOSS \citep{White:2010ed} and VIPERS (Guzzo et 
al. 2012, in preparation), are mapping significant volumes of the distant
universe and will produce measurements of 
$f\sigma_8$ covering the redshift range out to $z\sim 1.3$ . Even larger and deeper 
galaxy surveys are planned for the next ten years, both from ground
and space.  These are the proposed BigBOSS \citep{Schlegel:2011wb}
project using the refurbished KPNO and CTIO 4m telescopes, and, most
importantly, the ultimate redshift survey from space by the approved
ESA mission Euclid \citep{Laureijs2009} we already mentioned above.
Main aim of these projects is precisely the solution of the dark 
energy puzzle.
In particular, Euclid is a medium-size (M-class) mission of the ESA Cosmic
Vision programme,  which has recently been selected for 
implementation, with launch planned for 2019.  Euclid will perform
both a photometric survey in the visible and in three
near-infrared bands, to measure weak gravitational lensing by imaging $\sim 1.5$ billion galaxies, plus a spectroscopic slitless
survey of $\sim 65,000,000$ galaxies. Both surveys will be able to constrain
both the expansion and growth histories of the universe.  Their
combination makes Euclid a unique experiment, with superb precision and
optimal cross-control of systematic effects (see
\citealt{EditorialTeam:2011mu}).   In this paper we 
present the expected performances of the spectroscopic survey, as
described in the Euclid Definition Study
Report\footnote{http://www.euclid-ec.org}  \citep{EditorialTeam:2011mu}, in quantifying the growth
history and test for possible modifications of gravity as the origin
of cosmic acceleration.  

We shall first explore the forecasted constraints on the growth
rate both for the expected survey parameters, and for a pessimistic
galaxy density reduced by a factor of two.  
We will use the well-known parameterisation of the growth rate $f(z)
\simeq [\Omega_m (z)]^{\gamma}$ \citep{peebles, fry, lightman, Wang:1998gt}, where $\Omega_m = \rho_m/\rho_{\rm crit}$, $\rho_m$ is the matter energy density and $\rho_{\rm crit}$ is the critical density making the universe spatially flat,  to look at future constraints on the parameter
$\gamma$, which characterises the gravity model,  and test
degeneracies arising when making assumptions on the background
cosmology. 

We will then look at how constraints on growth are modified when survey specifications are changed, such as the total covered area and the corresponding galaxy number density (see discussion in Sec. \ref{sec.figure of merit})  and the redshift range covered by the survey. 
In order to do this we need a measure of the quality of our growth constraint. For this reason we introduce a new simple Figure of Merit (FoM), analogous to the dark energy FoM defined in the Dark Energy Task Force Report \citep{Albrecht:2006um}. We then show how this FoM depends on the already mentioned survey specifications.

Expressing $f$ as a function of $\gamma$ when considering sub-horizon matter perturbation is a well justified choice for GR cosmologies and for some modified gravity theories (e.g. the DGP model, a well-known five-dimensional model proposed in \citealt{Dvali:2000hr}  where the acceleration of the expansion is due to the leakage of gravity into an extra dimension) but it is not the only possibility. In some cases a constant $\gamma$ cannot describe the growth rate, as shown, for example, by \citet{DiPorto:2007ym} for a class of models of dark energy coupled to matter. Therefore, when optimising a survey configuration for the growth rate measurement, it is important to consider not only one parameterisation of $f$, in order not to bias our decisions.
The same was shown for the case of the equation of state parameterisation in \citet{Wang:2010gq}. To have a more general perspective, we take a second growth parameterisation and compare the results obtained. We choose the parameterisation proposed by \citet{Pogosian:2010tj}, \citet{Song:2010fg} which is physical, works at any scale and it is also able to reproduce the DGP growth rate.

Finally, we focus on the specific question ``will the Euclid spectroscopic survey be able to distinguish between GR and modified gravity models?'' To answer it, we use a Bayesian approach, following the work of \citet{Heavens:2007ka}. Here, a way to forecast the Bayesian evidence is proposed, based on Fisher statistics. In particular, we look at how much $\gamma$ of a modified gravity model has to differ from the GR value in order for our fiducial survey to be able to distinguish it.

Our work extends that of \citet{Wang:2010gq}, who considered using a Euclid-like survey to measure the dark energy equation of state, to further consider the growth rate of structure. It confirms the results of \citet{Simpson:2009zj} and \citet{Samushia:2010ki},  who specifically focused on cosmological model dependence.  Parallel work that appeared while this paper was in preparation includes that of \citet{DiPorto:2011jr}, who look at bias and time-dependent parameterisations of $\gamma$, while we specialise on different growth parameterisations and on the dependence on survey specifications,\cite{Belloso:2011ms}, who find an exact solution of $\gamma$ and \citet{Ziaeepour:2011bq}, who estimate constraints from wide surveys based on a new growth parameterisation for modified gravity and interacting dark energy models. A difference between our work and all the papers mentioned above is that our results are updated to the latest Euclid configuration \citep{EditorialTeam:2011mu}, which includes results from the most accurate simulations of the instrument.

The outline of the paper is the following.
In Sec. \ref{sec. redshift space distortions} we describe RSD as an observable to measure growth, while in Sec.~\ref{sec.Fisher_matrix_formalism} the Fisher matrix method used to forecast errors on growth and other cosmological parameters is outlined, and the fiducial cosmology is defined. Sec. \ref{sec.simulations} describes how the survey was modelled, while Sec.~\ref{sec.forecasts} shows our forecasts, and in the following Sec.~\ref{sec.figure of merit}
 we  define the growth Figure of Merit, which we apply to our forecasted survey data. In Sec. \ref{sec.bayesian evidence} we compute the forecasted Bayesian evidence and
we finally conclude in Sec.~\ref{sec.conclusions}. Appendix \ref{sec.growth of structures} contains a brief review of the theory of  the growth of linear small-scale perturbations and of growth parameterisations used in the literature.

\section{Measuring growth: redshift space distortions}\label{sec. redshift space distortions}

Our observable, containing both BAO and RSD, is the galaxy power spectrum.  We write the observed power spectrum $P_{\rm obs}$ as
\citep{Kaiser:1987qv, Se, Song:2008qt}
\bea \nonumber
P_{\rm obs}(k^\perp_{\rm r}, k^\parallel_{\rm r}) &=& \left[ \frac{D_A(z)_{\rm r}}{D_A(z)} \right]^2 \, \left[\frac{H(z)}{H(z)_{\rm r}}\right] \,\left(b \sigma_8  + f \sigma_8 \, \mu_{\rm r}^2 \right) ^2 \\
&& \left[\frac{P_{\rm matter}(k)}{\sigma_8^2}\right]_{z = 0} + P_{\rm shot} \,.\label{pk1}
\eea
where the subscript r indicates quantities in the reference cosmology chosen to compute the power spectrum, 
 $k^\perp_{\rm r} = k^\perp D_A(z)/D_A(z)_{\rm r}$, $ k^\parallel_{\rm r} = k^\parallel H(z)_{\rm r}/H(z)$ are the wave modes perpendicular and parallel to the line-of-sight, $k = \sqrt{ {k^\perp}^2 + {k^\parallel}^2 } $, $\mu = \textbf{k} \cdot \hat{\textbf{r}} /k = k^\parallel/k$  and $P_{\rm shot}$ is a scale-independent offset due to imperfect removal of shot-noise.  The term $\left[ D_A(z)_{\rm r}/D_A(z) \right]^2 \, \left[H(z)/H(z)_{\rm r}\right] $ represents  the distortion of the power spectrum due to the Alcock-Paczynski effect \citep{Alcock:1979mp}, which also has the impact of changing the true $(k,\,\mu)$ into reference $(k_{\rm r},\,\mu_{\rm r})$. 

Notice here that the term $\left[P_{\rm matter}(k)/\sigma_8^2\right]_{z = 0}$ is independent of the normalisation of the matter power spectrum.

We allow for the growth of structure to vary away from that of a LCDM model in two directions. First we allow the growth rate, which is well approximated by $f = \Omega_m^\gamma$ ($f= \Omega_m^\gamma +(\gamma -4/7)\Omega_k$ for curved space, \citealp{Gong:2009sp}) with $\gamma=0.545$ in LCDM models, to have $\gamma\ne0.545$. Second, we directly allow a modification of the standard equation for the time-like metric potential $\Psi$. This modification is described following \citet{Pogosian:2010tj, Song:2010fg} by a constant parameter  $\mu_s \ne 0$.
This choice and its motivation are discussed in more detail in Appendix A. We will use these parameters to model growth from observations of redshift-space distortions.

\section{Forecasting the errors}\label{sec.Fisher_matrix_formalism}
In this section we give details of the method used to estimate the expected minimum errors obtainable on $\gamma$ or $\mu_s$
and the other cosmological parameters from future redshift surveys, once the measurement error on the
observables is known.

To perform our forecasts we use the standard Fisher matrix approach which was introduced to forecast errors on $P(k)$ and derived parameters \citep{Tegmark:1997rp} and then adapted by \citet{Se} to the measurement of distances using the BAO position, avoiding
the ``noise'' of RSD.   A series of papers in the literature have
then used and discussed this technique to extract
the information on both expansion and growth (for example recently \citealt{Wang:2008, White:2008jy,
 Simpson:2009zj, Wang:2010gq, Samushia:2010ki, DiPorto:2011jr}).  Here we follow in
particular the approximations discussed in \citet{Samushia:2010ki}.

Briefly, the Fisher matrix is defined as (see e.g. \citealt{Bassett:2009uv}) $ F_{ij} = - \left\langle \partial^2 \ln \mathcal{L}/\partial p_i \partial p_j\right\rangle$,
where $\mathcal{L}$ is the likelihood function and $p_i$ are the
model parameters whose error one wishes to forecast.
It is possible to show that the Fisher matrix is the inverse of the covariance matrix, $F_{ij} = C^{-1}_{ij}$, in the (strong) assumption that the likelihood is a Gaussian function of the parameters and not only of the data.
The galaxy power spectrum Fisher matrix can be
approximated as \citep{Tegmark:1997rp}
\be \label{eq.fisher_Tegmark}
F_{ij} = \int_{k_{min}}^{k_{max}} \frac{\partial \ln P_{\rm obs}({\bf k})}{\partial p_i} \frac{\partial \ln P_{\rm obs}({\bf k})}{\partial p_j} V_{eff}({\bf k}) \frac{d {\bf k}^3}{2(2\pi)^3}
\ee
where $P_{\rm obs}({\bf k})$  is the galaxy power spectrum, Eq.~(\ref{pk1}), and its derivatives are computed in a chosen fiducial model, $V_{eff} =V_{surv} \left[1+1/(nP_{\rm obs}(k))\right]^{-2}$, $V_{surv}$ is the volume of the survey and $n$ is the galaxy number density.

Following \citet{Samushia:2010ki}, we neglect $P_{\rm shot}$,  since this term should only introduce negligible error.  We also neglect the dependence of $P_{\rm matter}(k)$ on $\Omega_b h^2$, $\Omega_m h^2$, $h$ and $n_s$. This means that we do not use the information coming from the shape of the matter power spectrum.
We finally multiply the integrand of the Fisher Matrix by a factor accounting for a possible error in redshift, $\exp[-\left(k_{\rm ref}\mu_{\rm ref}\, c\, \sigma_z/H_{\rm ref}\right)^2)]$, where $\sigma_z = 0.001 (1+z)$ is the standard deviation of the redshift error expected in Euclid.
 In each redshift bin centred at $z=z_i$ we compute a Fisher matrix \citep{Se} with parameters $p_j$
\be
p_j = \{ f(z_i) \sigma_8(z_i)\,,\, b(z_i)\sigma_8(z_i)\,,\,  \ln D_A(z_i) \,,\, \ln H(z_i)  \}\,,
\ee
where $p_j$ in different redshift bins are considered to be independent; then all the matrices corresponding to all $z_i$ are summed. From the total Fisher matrix it is possible to estimate the errors on each $p_j$ in each redshift bin, by marginalising over all other parameters. To obtain errors on cosmological parameters, we marginalise over $b(z_i)\sigma_8(z_i) $ and project the obtained matrix into the final cosmological parameter set.

Regarding the latter, we test three nested models: 1) a simple
quasi-LCDM with a constant $w$ fixed to $-0.95$ instead of $-1$ as in
LCDM, which we dub qLCDM\footnote{We take $w\neq -1$ to match the choice of \citet{EditorialTeam:2011mu}, motivated by the possibility of computing cosmological perturbations avoiding the $w=-1$ barrier.}, 2) a model where dark energy
has constant equation of state $w$, dubbed wCDM and 3) a model where
the equation of state of dark energy is allowed to vary, following the
evolution $w = w_0 + w_a(1-a)$, which we call CPL
\citep{Chevallier:2000qy, Linder:2002et}. For all these models we
consider both the flat and the curved space cases. The full set of
parameters, from which we pick the appropriate subset according
to the chosen model, is therefore 
\be 
q_j = \{ h\,,\,\, \Omega_m\,,\,\, \Omega_k\,,\,\, w_0 \,,\,\, w_a
\,,\,\, \gamma \,{\rm or}\,\mu_s\,,\,\, \sigma_8(z=0) \}\,, 
\ee
where $h = H_0/(100\, {\rm km}\, {\rm s}^{-1}{\rm Mpc}^{-1})$.  This is different from \cite{Samushia:2010ki} since they do not consider $\sigma_8(z=0)$ and their resulting errors will be consequently smaller with respect to this work, as we will see in Sec. \ref{sec.forecasts}.

Our fiducial model is, as in the Euclid Definition Report, a flat
constant $w$ cosmology with parameter values as the best-fit WMAP-7
\citep{Komatsu:2010fb} results  (except for $w$):  
\bea \nonumber
&& h = 0.703 \quad
\Omega_m  = 0.271 \quad
\Omega_{k} = 0.\quad
\sigma_8(z=0) = 0.809\quad \\
&& \nonumber w_0 = -0.95\quad
w_a = 0.\quad
\gamma = 0.545\,/ \mu_s = 0\\
&& \Omega_b  = 0.045 \quad
n_s = 0.966\quad \label{eq.fid.pars_in}
\eea
The fiducial matter power spectrum for this cosmology is computed
using CAMB\footnote{http://camb.info} \citep{Lewis:1999bs}. 

We assume a scale-independent bias, which is a good approximation for
large enough scales. As a fiducial bias, we take the bias function
derived by  \citet{Orsi:2009mj}  using a semi-analytical model of
galaxy formation.  For an analysis of how the bias can be constrained
and of the 
impact of assuming a biasing model on the estimates of the growth
factor we refer to the parallel work of \citet{DiPorto:2011jr}. Number
densities and biasing parameters are summarised in
Table~\ref{tab.density&bias_new}, together with the integration limits
in $k$. The latter correspond to scales $R$ such that $\sigma^2(R)=0.25$,
with an additional cut at $k_{\rm max}= 0.20 \,h\,Mpc^{-1}$. 

 In a separate paper, \citet{Bianchi:2012za} perform a detailed
study of statistical and systematic errors in RSD measurements, using
a large set of mock surveys built from numerical simulations. They
compare their results with predictions from the Fisher matrix code
used here, finding fairly good agreement when considering only linear scales.

\section{Modelling the Euclid Survey} \label{sec.simulations}
Euclid will cover  an area of $15,000\,{\rm deg}^2$ in both imaging and spectroscopy
\citep{EditorialTeam:2011mu}, measuring redshifts in the infrared band ($ 0.9-2 \;\mu m$) for $\sim 65$ Million galaxies, using a slitless spectrograph. With
this technique, the redshift measurement relies on the detection of
emission lines in the galaxy spectra, which in the chosen wavelength
range in the vast majority of cases will be the $H\alpha$ line,
redshifted to $0.7<z<2$.  

Intrinsic to the slitless technique is the impossibility to define {\it
  a priori} a survey flux limit (as normally done in a classical
slit survey, which is based on a well defined target sample selected
to a given magnitude or flux limit).  Spectra are intrinsically
confused by superpositions and the actual flux limit depends not only on
the nominal signal-to-noise reachable through a given exposure, but
also and fundamentally on the strategy devised as to resolve the
confusion among the different spectra.  In the case of Euclid, this is
achieved first by splitting the wavelength range into two
sub-exposures through  a ``blue'' and a ``red'' grisms, covering
respectively the wavelength ranges $0.9-1.4\; \mu m$ and $1.4-2.0\;\mu m$; this has the advantage of halving the length of the spectra on
the detector, thus reducing superpositions.  Secondly, the blue and
red exposures are in turn split into two sub-exposures, observed
rotating the field of view by $90^\circ$.  This makes for four
different exposures of a given field, which result in a sensible
treatment of spectral confusion.  To verify this and compute the
success rate of the survey (i.e. the fraction of correctly 
measured redshifts over the total number of spectra), it was necessary
to develop since the early stages of the project an advanced end-to-end
simulation pipeline.  An accurate description of these
simulations is beyond the scope of this paper, and we refer the reader to
the Euclid Definition Study Report \citep{EditorialTeam:2011mu}, and
to the specific paper (Garilli et al. 2012, in preparation).  Here we
recall only the main concepts, which are relevant for the present
analysis.  
 
\begin{figure}
\includegraphics[width=0.45\textwidth]{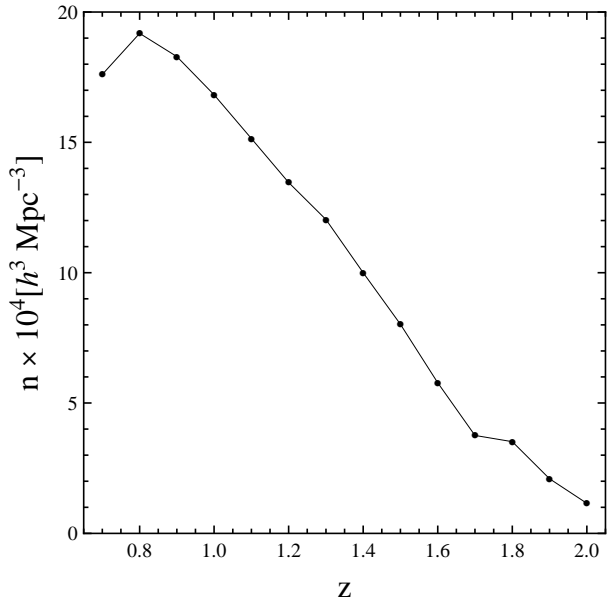}
\caption{Predicted mean number density of
  galaxies in each  redshift bin centred in $z$, expected from the baseline Euclid
  wide spectroscopic survey, given the instrumental and survey
  configurations and the estimated
  efficiency.}\label{fig.z-distribution} 
\end{figure}

The end-to-end spectroscopic simulations take into account Euclid's
instrumental (point-spread function, resolution
and instrumental background) and observational (exposure time,
astrophysical background) parameters, to build an artificial
``observation'' of a realistic galaxy field. 
The input data set is built starting from the COSMOS catalogue of photometric
redshifts \citep{Ilbert:2008hz}, which contains all relevant
information (coordinates, redshift, luminosity, spectral energy distribution, stellar formation rate, etc.).
Most importantly, the COSMOS depth and extensive wavelength coverage (over 30
spectral bands), allows us to assign a well-defined spectral type and
thus a realistic distribution of $H\alpha$ equivalent widths, down to
large distances.  It is also important that, being based on real
observations, the catalogue includes also a realistic clustering of
the sources. Stars are then added upon the galaxy catalogue.  The  
package aXeSIM3 is then used to generate the four 2D dispersed images of the field
and then extract the four 1D spectra for each target.  
Redshift are measured making use of RESS (Redshift Evaluation
from Slitless Spectroscopy), an automatic software running within
IRAF environment, which has been devised (Rossetti E. 2012, in
preparation) to reduce and analyse highly contaminated slitless spectra.
RESS' $z$ measure is currently implemented only for low $z$ galaxies
($0.7 < z < 2.0$) and its extension for high $z$ objects is in progress.
Redshift evaluation is based on the position of the $H \alpha$ line and
any other emission lines, when detected, for which a flux is also
measured.  A reliability flag for each measured redshift is then
obtained by further processing the spectra through the EZ redshift measurement
code \citep{Garilli}.  Comparison of the input and output catalogues
allows one to estimate the success rate of the survey in terms of {\it
  completeness} and {\it purity} as a function of
redshift and  $H\alpha$ flux (see Euclid Definition Study Report,
Fig. 6.10).  Rather than trusting the absolute redshift distribution
emerging from the simulated field, a more conservative choice is to use this output
as weight, to be applied to the most up-to-date predictions for the redshift distribution of 
$H\alpha$ emitters \citep{Geach:2008us}. This produces the expected
distribution of the number of galaxies with measured redshift in each
redshift bin. From this one can calculate the galaxy number density at each $z$, which is shown in Fig. \ref{fig.z-distribution} for our fiducial cosmology of Eq. (\ref{eq.fid.pars_in}). 
\begin{table}
\centering
\begin{tabular}{l@{\hspace{0.4cm}} l@{\hspace{0.4cm}} l@{\hspace{0.4cm}}  }
\\
\hline\hline\\
$z$ &  $b$ & $k_{\rm max} (h \, {\rm Mpc}^{-1})$\\
\hline\\
$0.7$ &	$1.083$& $0.1590$\\
$0.8$ &	$1.125$& $0.1691$\\
$0.9$ &	$1.104$& $0.1804$\\
$1.0$ & 	$1.126$& $0.1917$\\
$1.1$ & 	$1.208$& $0.1958$\\
$1.2$ & 	$1.243$& $0.2000$\\
$1.3$ & 	$1.282$& $0.2000$\\
$1.4$ & 	$1.292$& $0.2000$\\
$1.5$ & 	$1.363$& $0.2000$\\
$1.6$ & 	$1.497$& $0.2000$\\
$1.7$ & 	$1.486$& $0.2000$\\
$1.8$ & 	$1.491$& $0.2000$\\
$1.9$ & 	$1.573$& $0.2000$\\
$2.0$ & 	$1.568$& $0.2000$\\
\hline\hline
\end{tabular}\caption{Galaxy biasing
  parameter $b$ and $k_{\rm max}$ of integration for each redshift bin
  centred in $z$ for the Euclid spectroscopic survey baseline
  configuration, having an observed area of $15,000 \,
  {\rm deg}^2$}\label{tab.density&bias_new} 
\end{table}

\section{Standard predictions for Euclid}\label{sec.forecasts}
\begin{figure*}
\includegraphics[width=0.8\textwidth]{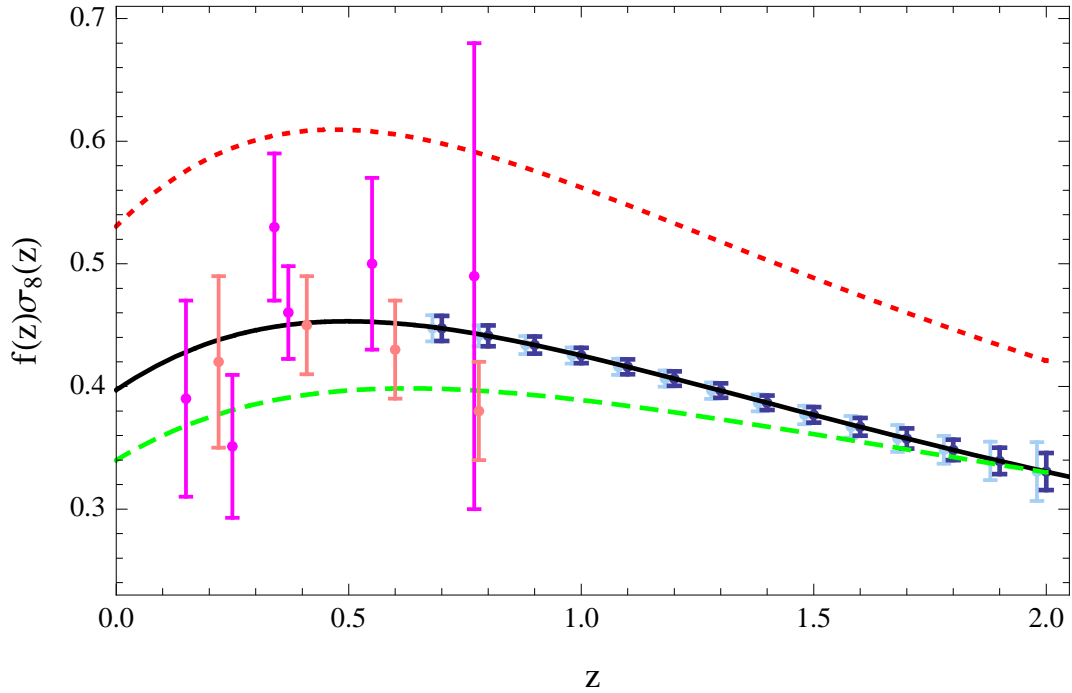}
\caption{Fisher matrix forecasts of the errors expected on the growth
 rate (dark-blue error bars), expressed through the bias-free combination $f(z) \sigma_8
 (z)$, obtainable from the Euclid redshift survey through
 the combination of amplitude and redshift-space anisotropy of galaxy
clustering. The light-blue error bars (shown with a slight offset in redshift for visualisation purposes) represent the case of a  galaxy density reduced by a factor of two with respect to that forecasted for the galaxies observed by Euclid \citep{Geach:2008us}. The solid black line represents the fiducial $f\,\sigma_8$, computed for the cosmology shown in Eq.~(\ref{eq.fid.pars_in}). The dashed green line shows the growth of a flat DGP model (calculated by numerical integration of the corresponding equation for $f(z)$). The red dotted line represents $f\,\sigma_8$ of a coupled model with coupling parameter $\beta_c = 0.2$. All models are computed for $\Omega_{m} = 0.271$ and for the same $\sigma_8(z_{CMB})$ as for the fiducial model. In the same plot we also show measurements of $f\,\sigma_8$ from past surveys (magenta error bars) and the recent WiggleZ survey (pink error bars), see explanation in the text.}\label{fig.fsig8}
\label{fig.errors.growth}
\end{figure*}

For our computations here, we split the Euclid predicted
redshift distribution over the range $0.7<z<2$,
into 14 bins with $\Delta z = 0.1$.  Using the predicted galaxy number
density in each bin shown in Fig. \ref{fig.z-distribution}, we obtain the error on our observable,
the power spectrum, and estimate the resulting precision on the
measurement of $f\, \sigma_8$ after marginalisation over the other parameters.
We plot errors on $f\, \sigma_8$  in Fig.~\ref{fig.errors.growth} (dark blue error bars), where we also show for comparison current measurements of $f\,\sigma_8$ (light pink and magenta error bars) and the pessimistic case of observing only half the number of galaxies forecasted in \citet{Geach:2008us} (light blue error bars), as the authors themselves claim that their counts may be wrong by a factor of 2.

Current measurements shown in Fig. \ref{fig.fsig8} are listed in Table \ref{tab.current-data}. The values of $f\,\sigma_8$ are computed in the case of \citet{Guzzo:2008ac} and \citet{Hw} by using the value of $f/b$ given by the authors and computing $b \sigma_8$ from $b$ and the reference cosmology they adopt for the computation of $b$ (or of \citealt{Lahav:2001sg} in the case of \citealt{Hw}); in the case of  \citet{Ro} $b \sigma_8$ was computed using the expression\footnote{This formula actually gives us the non-linear $b \sigma_8$, since we have used the non linear estimate of $\xi$ of \citet{Ro} to compute it. What we needed to obtain the {\emph{linear}} $f\sigma_8$ would be the linear $b\sigma_8$, but we do not have it. Therefore our estimate of $f\sigma_8$ for the \citet{Ro} datapoint might be $5-10\%$ higher than it should.}  \citep{Zehavi:2004zn}, ${(b \sigma_8)}^2 =\int^2_0 dy\,y^2\, \xi(8y)\,(3-9y/4+3y^3/16)$.  
\citet{CGA} indicate directly their value of $b \sigma_8$, while \cite{Blake:2011rj} and \citet{Samushia:2011cs}  compute directly $f \sigma_8$. Error bars are obtained through the error propagation formula for uncorrelated data, when not directly specified in the papers. 

\begin{table}
\centering
\begin{tabular}{l@{\hspace{0.4cm}} l@{\hspace{0.4cm}} l@{\hspace{0.4cm}} l@{\hspace{0.4cm}}  }
\\
\hline\hline\\
survey &  reference paper & $z$ & $f \sigma_8$\\
\hline
\hline\\
VVDS F22 	& \citet{Guzzo:2008ac}		&	$0.77$	& $0.49\pm 0.19$\\
wide			&						&			&			    \\
\hline
2SLAQ		& \citet{Ro} 				&	$0.55$	& $0.50 \pm 0.07$\\
galaxy		&						&			&			     \\
\hline
SDSS LRG	& \citet{CGA}			 	&	$0.34$	& $0.53\pm 0.07$\\
			& \citet{Samushia:2011cs}	&	$0.25$	& $0.35\pm 0.06$\\
			& 						&	$0.37$	& $0.46\pm 0.04$\\
\hline
2dFGRS		& \citet{Hw}				&	$0.15$	& $0.39 \pm 0.08$\\
\hline
WiggleZ		&  \cite{Blake:2011rj} 		&	$0.22$	& $0.49 \pm 0.07$\\
			&						&	$0.41$	& $0.45 \pm 0.04$\\
			& 						&	$0.6$	& $0.43 \pm 0.04$\\
			& 						&	$0.78$	& $0.78 \pm 0.04$\\
\hline\hline
\end{tabular}\caption{Current measurements of $f \sigma_8$}\label{tab.current-data}
\end{table}

Together with the (solid black) curve representing our fiducial $f\,\sigma_8$, we also show for comparison a (dashed green) line for flat DGP,  (calculated by numerical integration of the corresponding equation for $f$)  and a (dotted red) line for the coupled model of \citet{DiPorto:2011jr}, computed using the parameterisation of \citet{DiPorto:2007ym} with a coupling $\beta_c = 0.2$ (both with $\Omega_{m} = 0.271$ and the same $\sigma_8(z_{CMB})$ of our fiducial model).

We notice that we reach accuracies between $1.3\%$ and $4.4\%$ in the measurement of $f \,\sigma_8$ depending on the redshift bin, where the highest precision is reached for redshifts $z\simeq 1.0$.

\subsection{Comparison to other surveys}
\begin{figure}
\includegraphics[height=0.28\textheight]{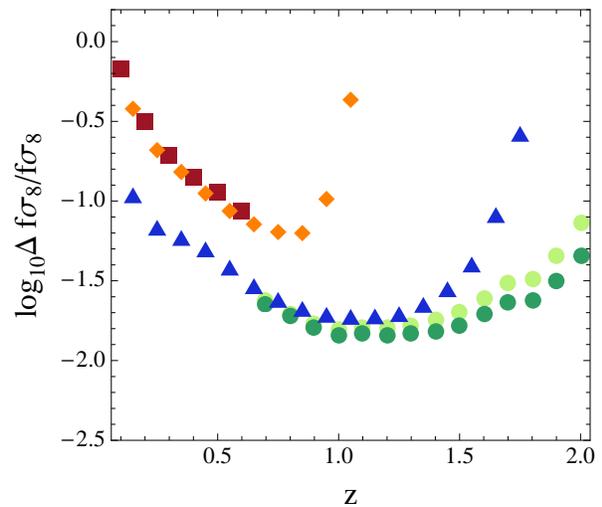}
\caption{Relative error on $f\,\sigma_8$ of Euclid (dark-green circles, light-green circles for the pessimistic case of half the galaxy number density), BOSS (dark-red squares), BigBOSS ELGs (blue triangles) and LRGs (orange diamonds).}
\label{fig.relative-errors-comparison}
\end{figure}
Together with Euclid, other ongoing and future surveys will constrain cosmology by measuring $f\sigma_8$. Here we compare the relative errors on $f \sigma_8$ obtained using different spectroscopic galaxy redshift surveys. 
In particular, we consider the BOSS survey\footnote{http://cosmology.lbl.gov/BOSS/} (see \citealt{Schlegel:2009hj}) and the BigBOSS\footnote{http://bigboss.lbl.gov/} Emission Line Galaxies (ELGs) and Luminous Red Galaxies (LRGs)\footnote{We thank the BigBOSS consortium for providing their latest yet unpublished estimate of their expected galaxy densities, which we used in creating this plot.}.
Regarding the fiducial bias, we use the forecasts by \citet{Orsi:2009mj} for BigBOSS ELGs. We use $b = 2 \, G(0)/G(z)$ (where $G(z)$ is the standard linear growth rate)  for BOSS and BigBOSS LRGs (see \cite{Reid:2009xm}).
Table \ref{tab.other-surveys} summarises the main characteristics of these surveys.

\begin{table*}
\centering
\begin{tabular}{l@{\hspace{0.4cm}} c@{\hspace{0.4cm}} r@{\hspace{0.4cm}}  r@{\hspace{0.4cm}} r@{\hspace{0.4cm}}  }
\\
\hline\hline\\
survey			&  redshift range	& area $[{\rm deg}^2]$	& $ n$ $[h^3\,{\rm Mpc}^{-3}]$			& bias				\\
\hline
\hline\\
BOSS LRG		& $0.05<z<0.65$	&$10,000$		&	$3 \times 10^{-4}$					& $2.0\,G(0)/G(z)$		\\
\hline
BigBOSS LRG 		& $0.1<z<1.1$		&$14,000$		&	unpublished (see footnote 7)		& $2.0\,G(0)/G(z)$		\\
\hline
BigBOSS ELG 		& $0.1<z<1.8$		&$14,000$		&	unpublished (see footnote 7)		& see \cite{Orsi:2009mj}	\\
\hline\hline
\end{tabular}\caption{Future and ongoing galaxy redshift surveys and their main properties}\label{tab.other-surveys}
\end{table*}

The results are shown in Fig. \ref{fig.relative-errors-comparison}. We first notice that Euclid (represented by dark-green circles) will obtain the most precise measurements of growth, even in the pessimistic situation of detecting only half the galaxies (light-green circles). In redshift coverage it will be perfectly complementary to BOSS. The partial overlap with BigBOSS, whose ELG sample will reach similar errors up to $z \sim 1.4$, will allow for interesting and useful independent measurements and cross-checks.
\begin{figure*}
\begin{center}
\includegraphics[height=0.3\textheight]{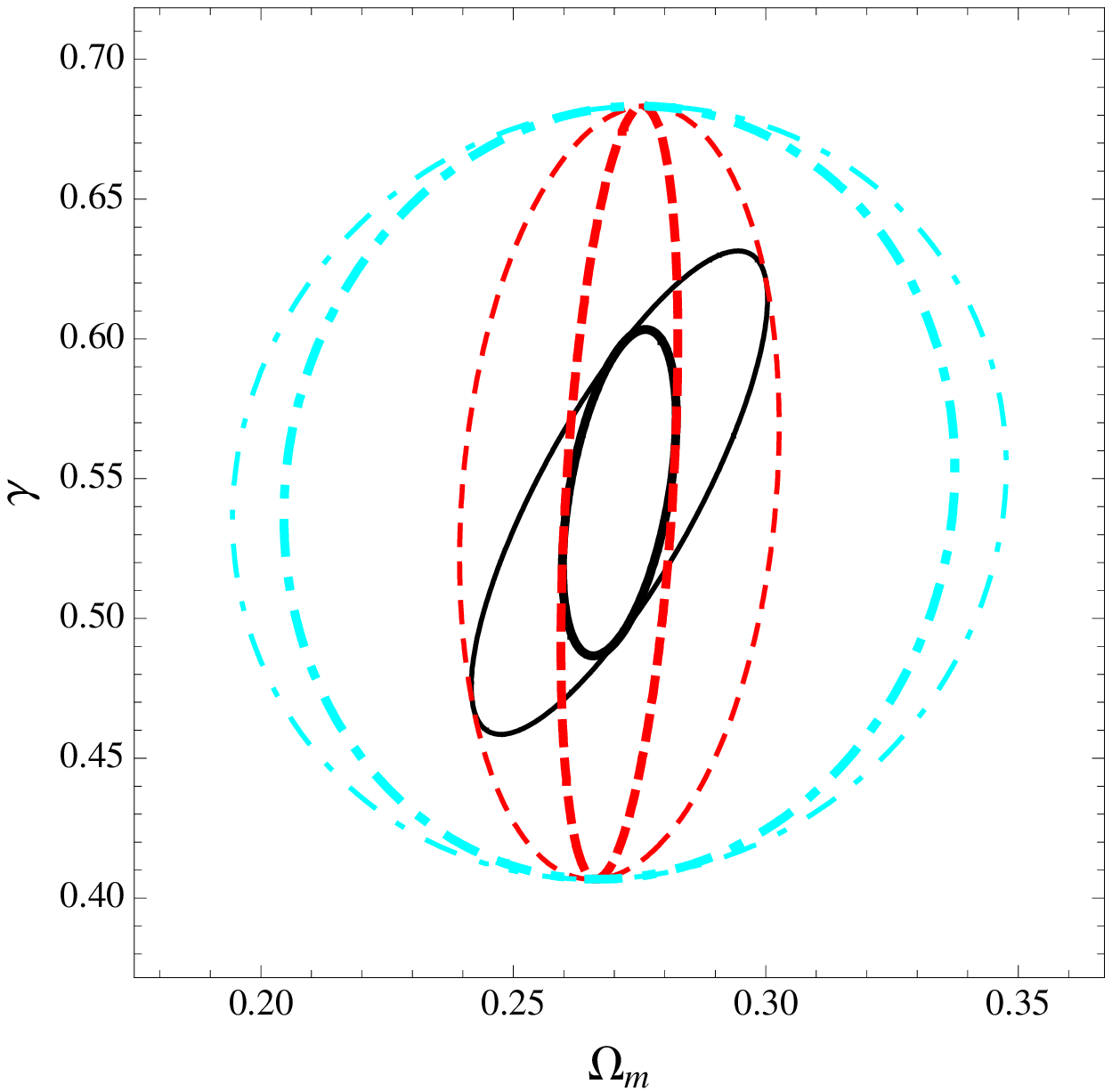} \quad
\includegraphics[height=0.3\textheight]{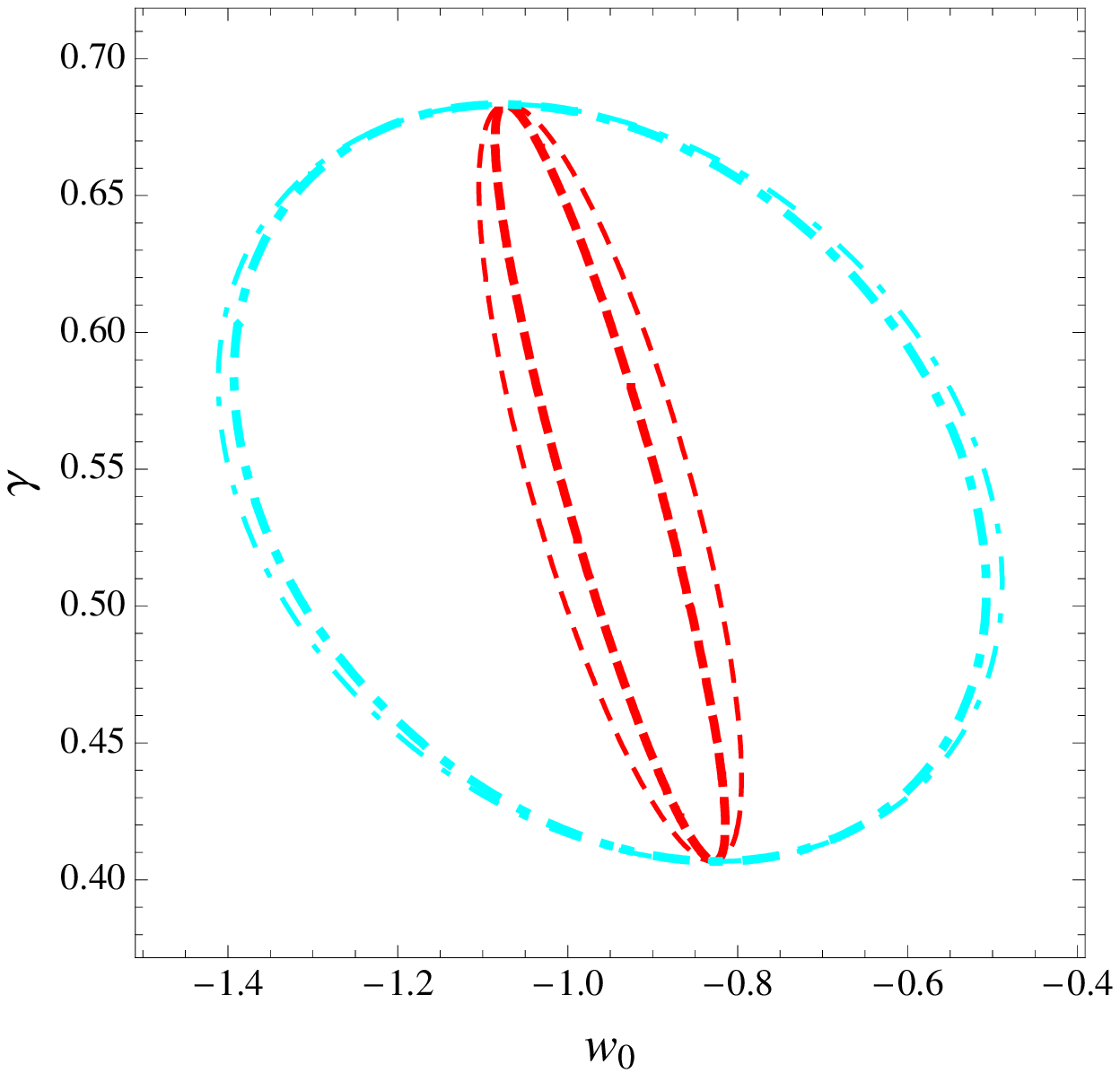} 
\caption{Forecasted errors on cosmological parameters using the Euclid spectroscopic survey.  Solid (black) lines correspond to qLCDM, dashed (red) lines to wCDM and dot-dashed (cyan) lines to CPL. Thick lines represent models with $\Omega_k = 0$, while thin lines represent curved models.  {\bfseries Left plot}: marginalised errors on $\Omega_m$ and $\gamma$. {\bfseries Right plot}: marginalised errors on $w_0$ and $\gamma$.}
\label{fig.fisher-growth}
\end{center}
\end{figure*}

\subsection{Cosmological parameters}

\begin{figure}
\begin{center}
\includegraphics[height=0.4\textwidth]{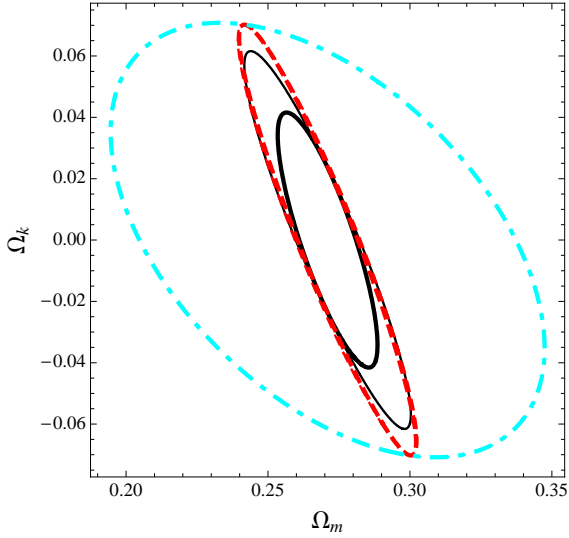}
\caption{Forecasted marginalised errors on $\Omega_m$ and $\Omega_k$ using the Euclid spectroscopic survey. Solid (black) lines correspond to qLCDM, dashed (red) lines to wCDM and dot-dashed (cyan) lines to CPL. Thick lines correspond to models where $\gamma = 0.545$ (its GR value) while thin lines represent models where $\gamma$ is allowed to assume other (constant) values. Note that for wCDM and CPL the thick and thin lines are one on top of each other: fixing $\gamma$ does not have any impact on constraints on $\Omega_m$ and $\Omega_k$.}
\label{fig.fisher-bg_1a}
\end{center}
\end{figure}

\begin{figure*}
\begin{center}
\includegraphics[height=0.3\textheight]{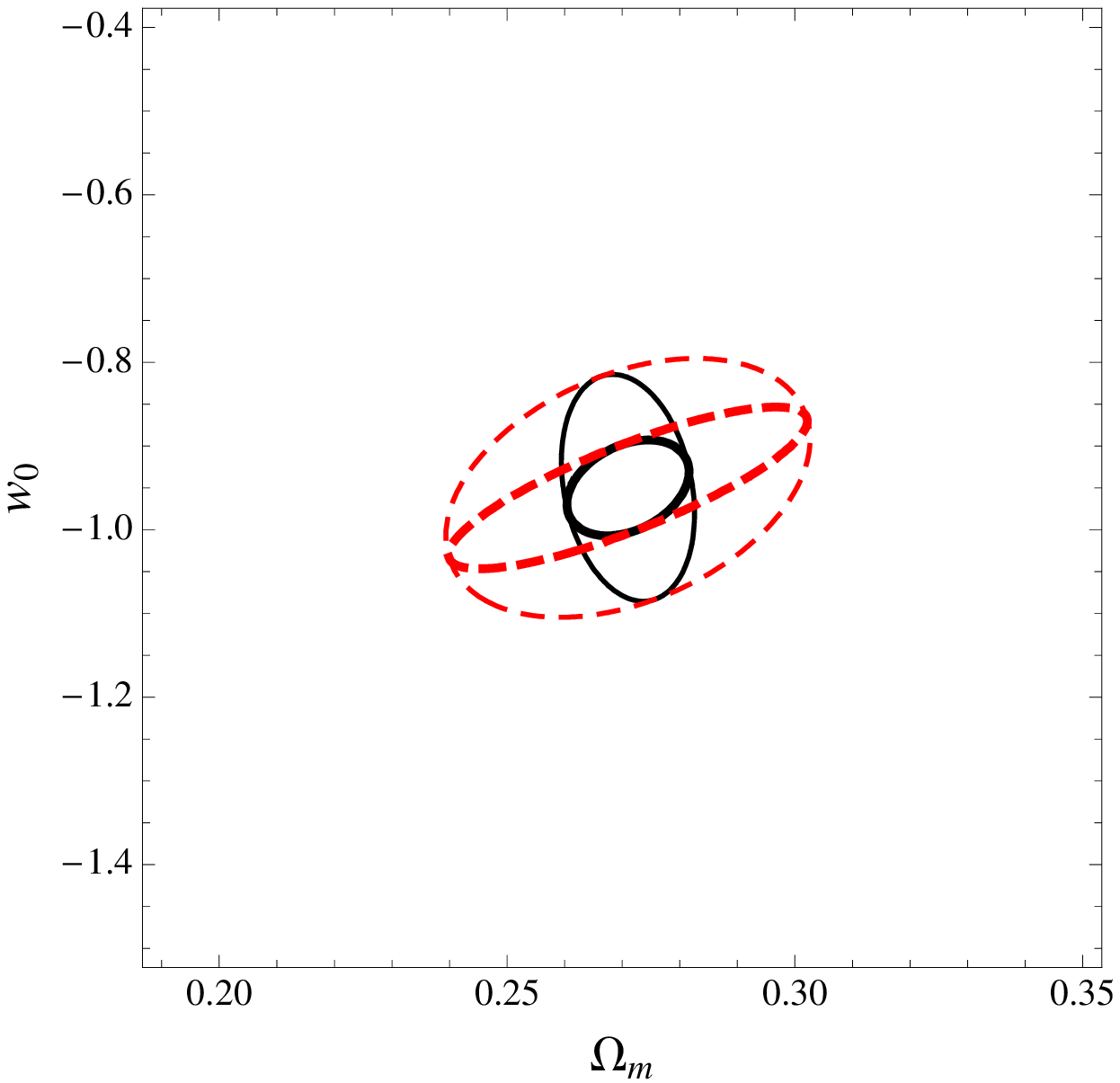}\quad
\includegraphics[height=0.3\textheight]{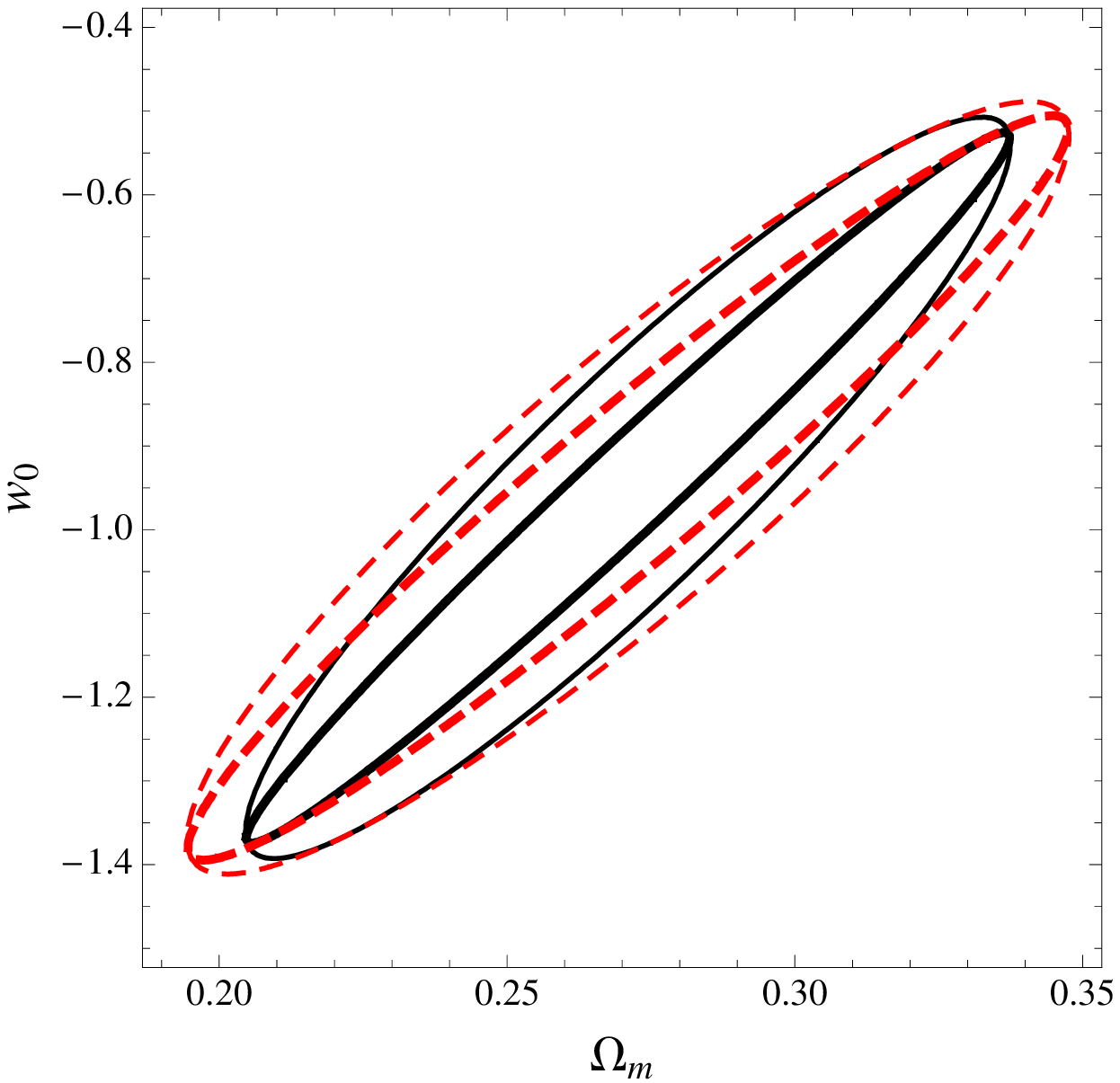}
\caption{Forecasted errors on $\Omega_m$ and $w_0$ with the Euclid spectroscopic survey.  Solid (black) lines correspond to models with $\Omega_k = 0$  and dashed (red) lines to   curved models. Thick lines refer to models  where $\gamma$ is fixed to the GR value,  while thin lines represent models where $\gamma$ is allowed to vary. {\bfseries Left panel}: wCDM model. {\bfseries Right panel}: CPL model }
\label{fig.fisher-bg_2}
\end{center}
\end{figure*}

\begin{figure}
\begin{center}
\includegraphics[height=0.3\textheight]{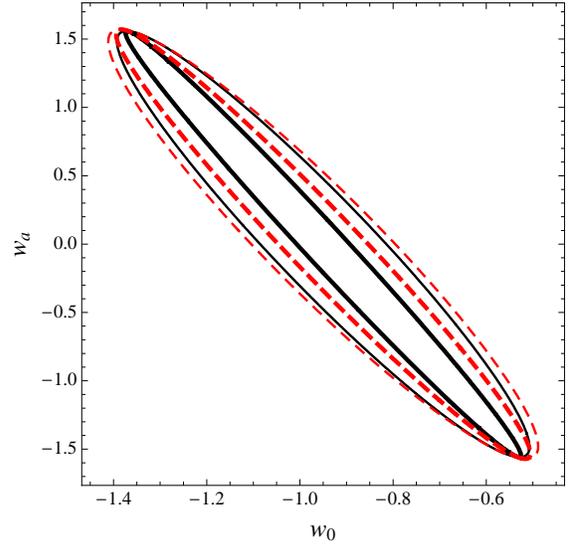}
\caption{Forecasted errors on the CPL parameters $w_0$ and $w_a$ using the Euclid spectroscopic survey. Solid (black) lines correspond to models with $\Omega_k = 0$, and dashed (red) lines to  curved models. Thick lines refer to models  where $\gamma = 0.545$,  while thin lines  represent models where $\gamma$ is allowed to take other (constant) values.}
\label{fig.fisher-bg_1b}
\end{center}
\end{figure}

We next look at how the errors on $ f\,\sigma_8$, $H$ and $D_A$ project into errors on cosmological parameters. We only show results for $\gamma$ and briefly mention results for $\mu_s$ (see Appendix \ref{app.mu_s-forecasts} for more detailed error forecasts), while in the next section we will compare the two parameterisations $\gamma$ and $\mu_s$. In order to understand the model-dependence of these forecasts, we use the nested cosmologies described in Sec. \ref{sec. redshift space distortions}: the more complicated model is a generalisation of the less complicated one, and the generalisation consists in the addition of one extra parameter. We also study the influence  in the estimate of background parameters of assuming GR or allowing for a different constant $\gamma$. Results are shown in Figs. \ref{fig.fisher-growth}-\ref{fig.fisher-bg_1b}.

From Fig.~\ref{fig.fisher-growth}~and~\ref{fig.fisher-bg_1a}, and by comparing left with right panel of Fig.~\ref{fig.fisher-bg_2} it is clear that what affects most strongly the error estimate, by enlarging the ellipses and even changing the degeneracy direction of parameters, is the assumption of different dark energy models. The assumption of zero curvature also has an influence. The less parameters the cosmological model possesses, the stronger is the influence of fixing $\Omega_k$, as can be seen from Fig.~\ref{fig.fisher-growth} (compare thick and thin lines having the same line style/colour)  and Fig.~\ref{fig.fisher-bg_2} (compare solid and dashed lines with same thickness). $\Omega_k = 0$ affects more strongly the measure of background parameters, and only indirectly constraints on $\gamma$ (note the thickening of error ellipses in Fig.~\ref{fig.fisher-growth} in the direction of background parameters $\Omega_m$ and $w_0$, respectively).

 As regards the $\mu$ parameterisation, the error on $\mu_s$ is forecasted to be between $\sim 0.5$ and $\sim 2$, depending on the dark energy model. The assumptions on $w$ and those on $\Omega_k$ affect more strongly constraints on this growth parameter, in the case of the $\mu_s-\Omega_m$ error ellipses.

The assumption of GR has a different influence on different pairs of parameters. As regards $\Omega_m-\Omega_k$, we can see from Fig.~\ref{fig.fisher-bg_1a} that fixing $\gamma$ reduces only the error for qLCDM, while errors on wCDM and CPL are unaffected. 
Errors on $w_0-w_a$ are very weakly affected by the assumption of GR and also of zero curvature (see Fig.~\ref{fig.fisher-bg_1b}), so that their determination is rather robust. 
 This conclusion agrees with what obtained in \cite{Samushia:2010ki} (see their Figs. 4 (a) and 5 (a) of the final published version), where the small differences are due to the presence of an extra parameter in our analysis over which we marginalise, namely $\sigma_8(z=0)$ and to the different survey specifications. The different degeneration direction of $\Omega_m-\Omega_k$ is due to the wrong sign convention being adopted in figure 5 of \cite{Samushia:2010ki} which had the effect of inverting the $\Omega_k$ axis. We have checked that our code gives exactly the same results as \cite{Samushia:2010ki} if the parameter $\sigma_8(z=0)$ is fixed and the survey specifications are identical. 
The pair of parameters which is most affected by the choice of fixing $\gamma$ is $\Omega_m-w_0$, more strongly in the case of wCDM, as can be seen in Fig. \ref{fig.fisher-bg_2}.

\subsection{Adding Planck}

\begin{figure*}
\begin{center}
\includegraphics[height=0.3\textheight]{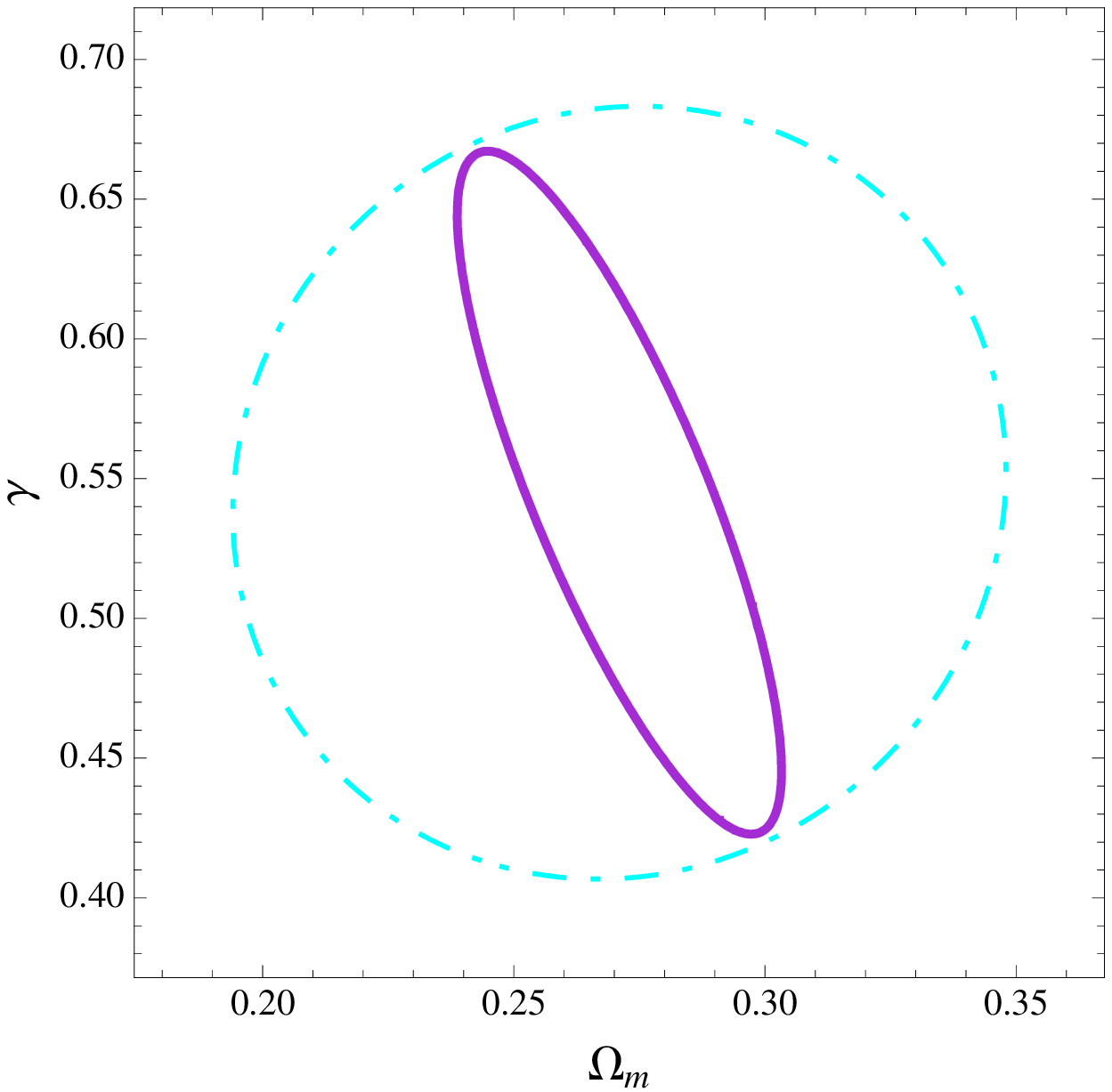} \quad
\includegraphics[height=0.3\textheight]{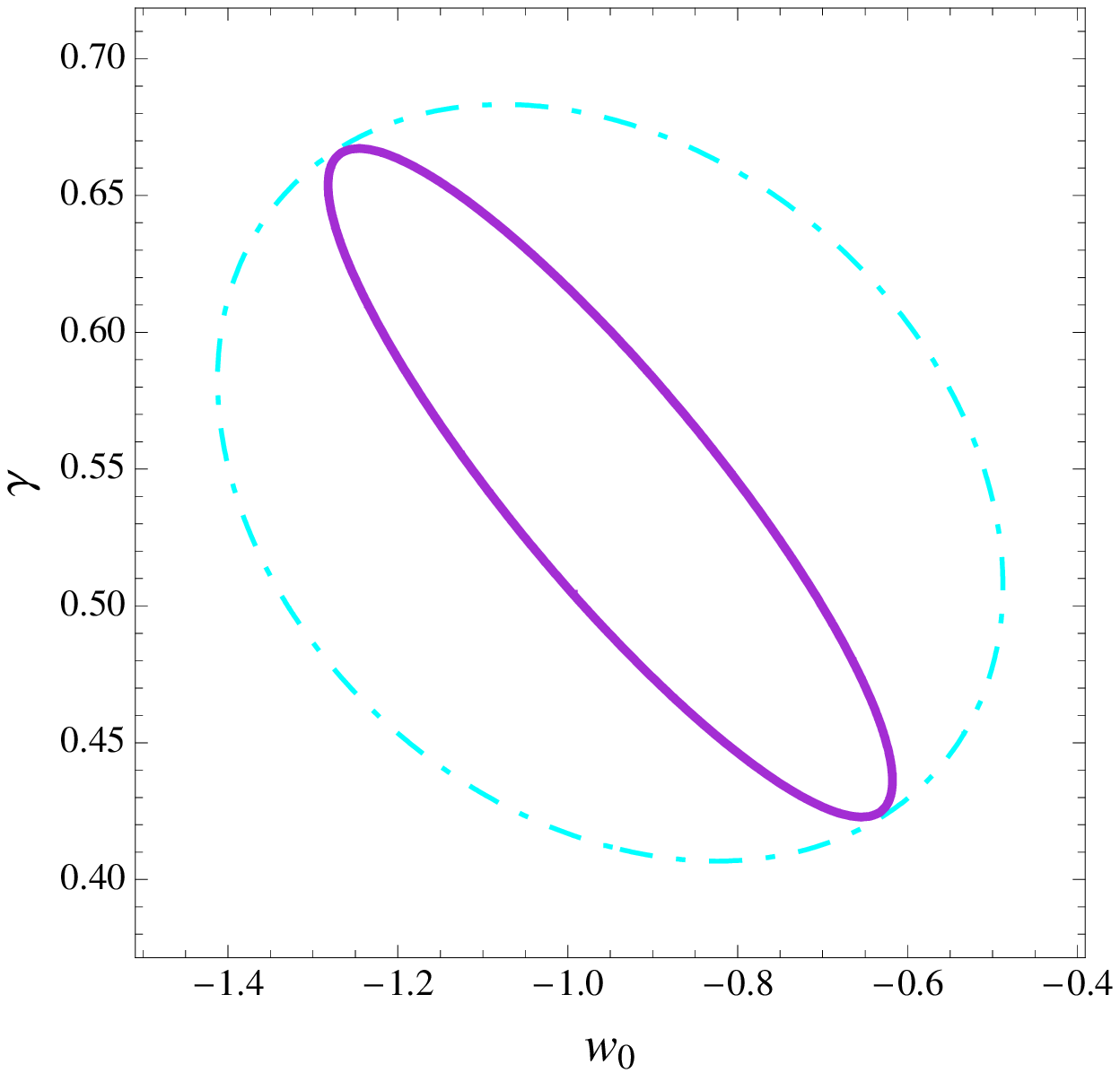}
\caption{Comparison of marginalised errors on curved CPL, obtained using Euclid only and joint Euclid and Planck data. Dot-dashed (cyan) contours correspond to error ellipses using Euclid data only, while solid (purple) contours show joint constraints from Euclid and Planck. {\bfseries Left plot}: marginalised errors on $\Omega_m$ and $\gamma$. {\bfseries Right plot}: marginalised errors on $w_0$ and $\gamma$.}
\label{fig.Planck_CPL_gamma}
\end{center}
\end{figure*}

\begin{figure}
\begin{center}
\includegraphics[height=0.3\textheight]{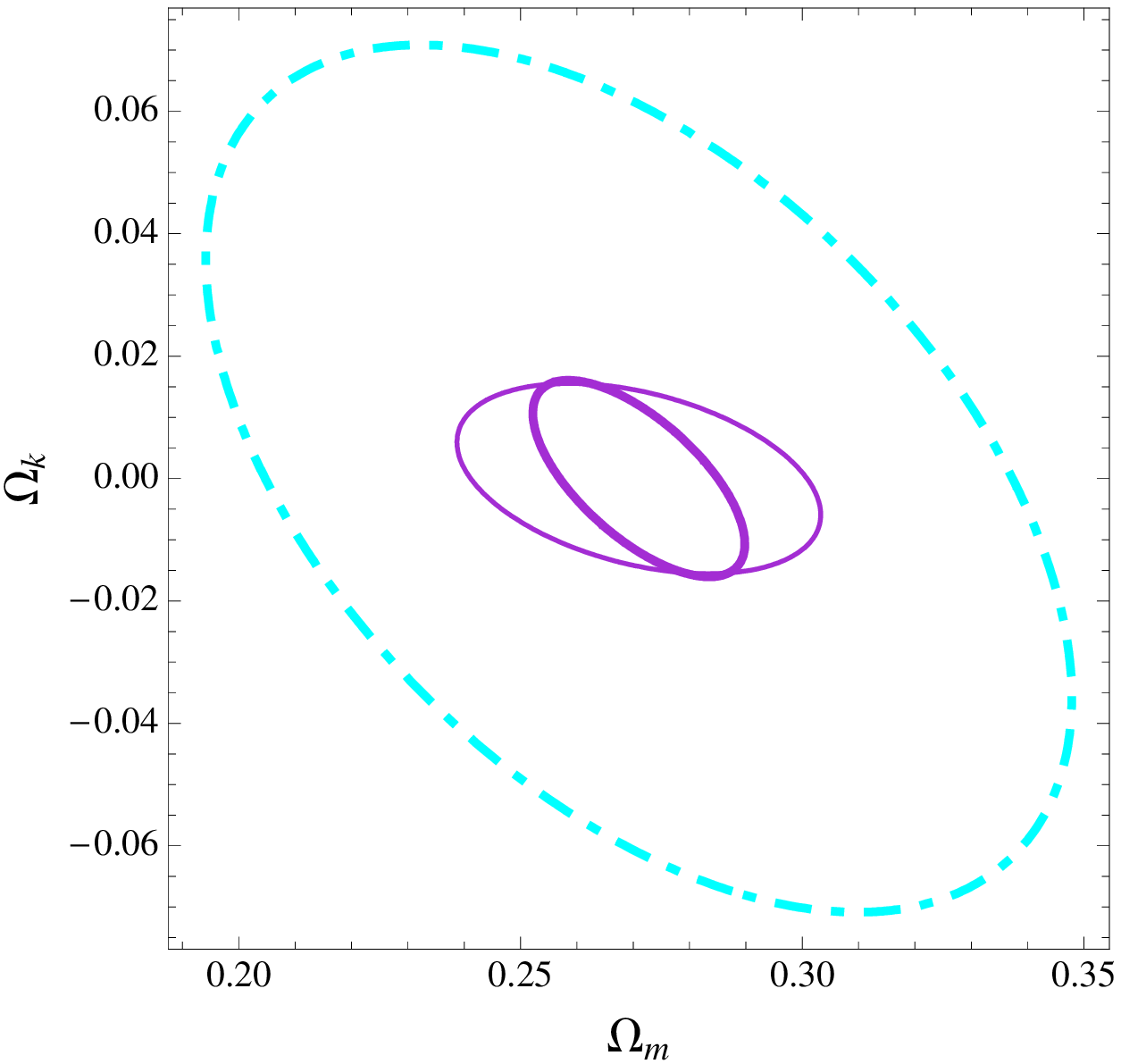}
\caption{Comparison of marginalised errors on background parameters $\Omega_m $ and $\Omega_k$ of a curved CPL model, obtained using Euclid and joint Euclid and Planck data. Dot-dashed (cyan) contours correspond to error ellipses using Euclid data only, while solid (purple) contours show joint constraints from Euclid and Planck.  Thick lines refer to models where $\gamma = 0.545$, while thin lines correspond to models where $\gamma$ is allowed to vary. Thin and thick cyan lines are superimposed because the CPL constraints from Euclid only data do not change when fixing $\gamma$.}
\label{fig.Planck_CPL_bg-1}
\end{center}
\end{figure}

\begin{figure}
\begin{center}
\includegraphics[height=0.3\textheight]{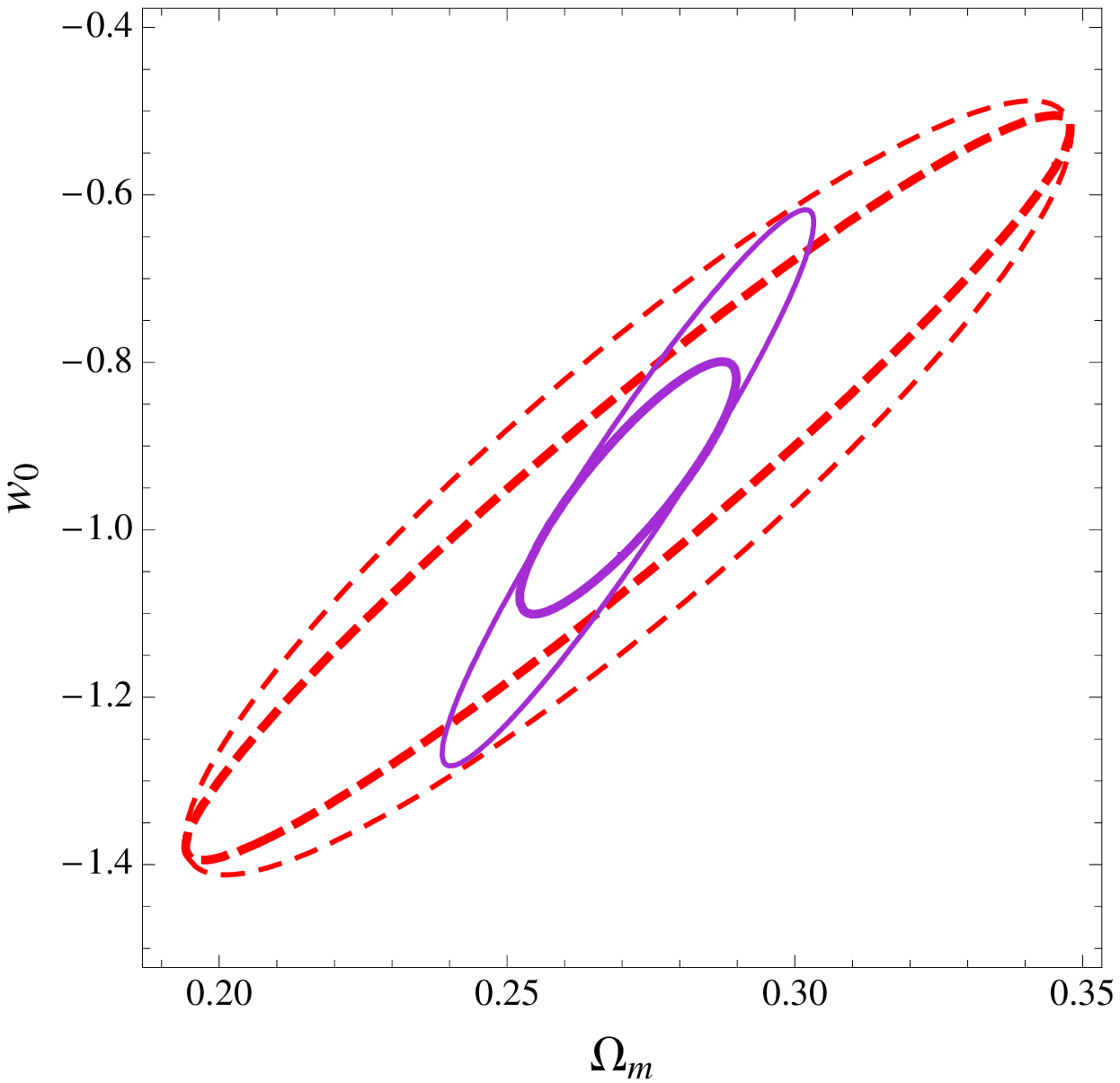}
\caption{Comparison of marginalised errors on background parameters $\Omega_m $ and $w_0$ of a curved CPL model, obtained using Euclid and joint Euclid and Planck data. Dashed (red) contours correspond to error ellipses using Euclid data only, while solid (purple) contours show joint constraints from Euclid and Planck.  Thick lines refer to models where $\gamma = 0.545$, while thin lines correspond to models where $\gamma$ is allowed to vary.}
\label{fig.Planck_CPL_bg-2}
\end{center}
\end{figure}

\begin{figure}
\begin{center}
\includegraphics[height=0.3\textheight]{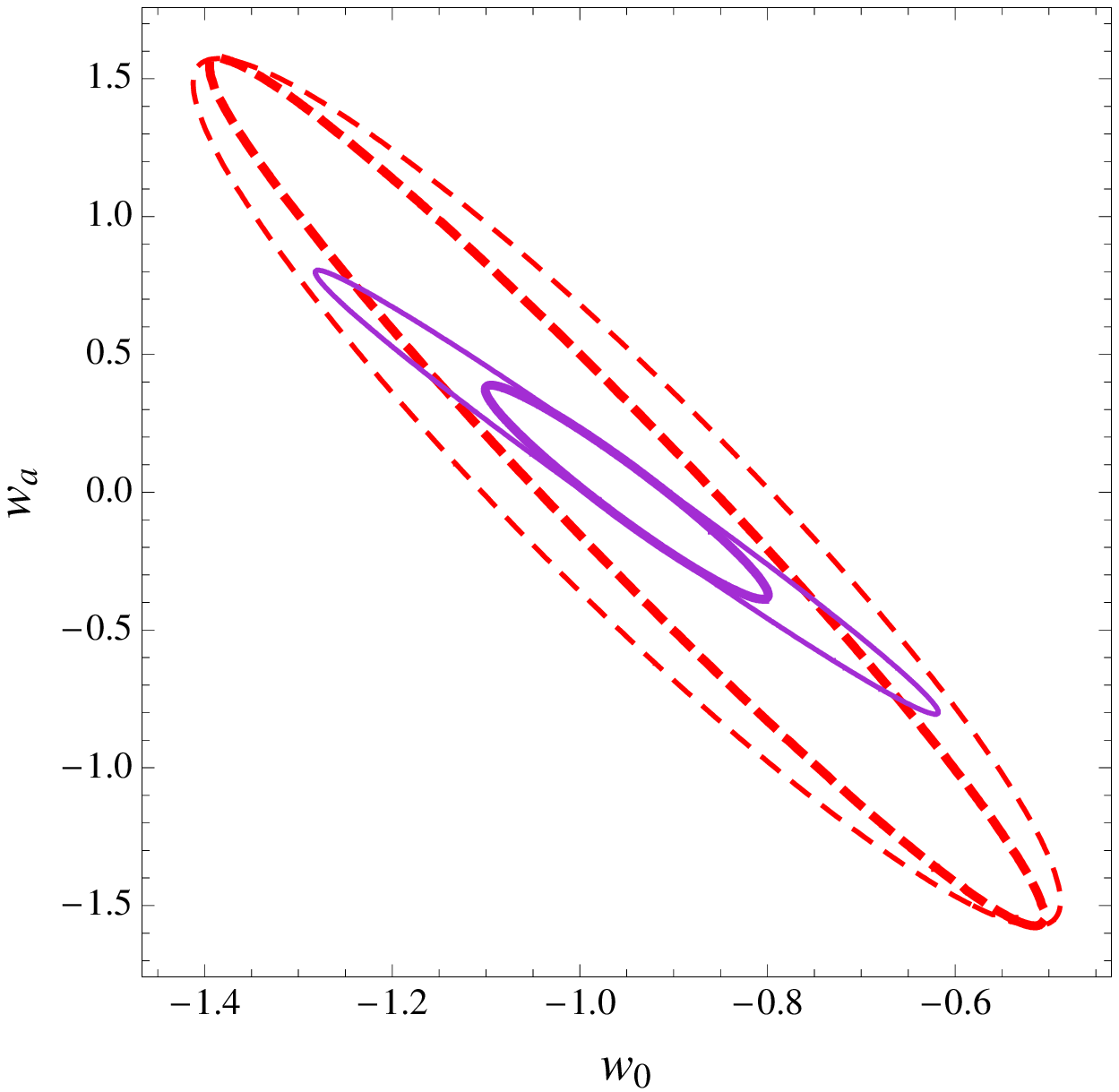}
\caption{Comparison of marginalised errors on the curved CPL background parameters $w_0 $ and $w_a$, obtained using Euclid and joint Euclid and Planck data. Colours and line styles are the same as in Fig.~\ref{fig.Planck_CPL_bg-2}.}
\label{fig.Planck_CPL_bg-3}
\end{center}
\end{figure}

We then quantify the impact of adding Planck constraints to the Euclid spectroscopic survey constraints, to check how much errors reduce. The Planck satellite (part of the Cosmic Vision programme by the European Space Agency) has been launched in 2009 and it is presently operating \citep{Planck_pre-launch}.  Its results on cosmology will be made public in the next years and will be the state-of-the-art data on CMB temperature and polarisation. It is natural and convenient to combine these data to Euclid galaxy survey data, first of all because these probes are highly complementary, observing the early and late universe respectively, and secondly because of the presence of the BAO feature in both datasets. As in \citet{Samushia:2010ki}, we utilise the Dark Energy Task Force Planck Fisher, computed for 8 parameters: $h$, $\Omega_m$, $\Omega_k$, $w_0$, $w_a$, $\sigma_8(z=0)$, $n_s$ and $\Omega_b$. Since this  Fisher matrix is computed assuming GR, to generalise it for arbitrary $\gamma$ in order to sum it to our galaxy survey Fisher matrix, we use the same method as in \citet{Samushia:2010ki} (see their Appendix B for details). In Figs.~\ref{fig.Planck_CPL_gamma}-\ref{fig.Planck_CPL_bg-3} we compare errors with (solid purple lines) and without (red-dashed or cyan dot-dashed lines) adding Planck, for a CPL dark energy.

As can be seen from  Fig.~\ref{fig.Planck_CPL_gamma}, Planck  noticeably improves marginalised errors on $\gamma$, obviously by constraining the background parameters. The error most reduced by Planck is, as one expects, that on $\Omega_m-\Omega_k$ (see Fig.~\ref{fig.Planck_CPL_bg-1}), quite independently of the assumptions on  growth (compare thick lines-corresponding to assuming GR, to thin lines--where $\gamma$ is not fixed). As regards constraints on the alternative $\mu$ parameterisation, here the impact  of adding Planck data on combined $\mu_s-\Omega_m$ and $\mu_s-w_0$ constraints is stronger than for $\gamma$: the projected error on $\mu_s$ is reduced from $\sim 2$ to $\sim 0.7$. This is shown in more detail in Appendix \ref{app.mu_s-forecasts}.
Joint Euclid-Planck constraints on $\Omega_m-w_0$ (Fig.~\ref{fig.Planck_CPL_bg-2}) and $w_0-w_a$ (Fig.~\ref{fig.Planck_CPL_bg-3}) depend more strongly on the assumption on $\gamma$, but are in any case a decisive improvement with respect to Euclid-only constraints.  This is again consistent with what obtained by \cite{Samushia:2010ki}, who find slightly tighter constraints for the reason explained above.

\section{A Figure of Merit for growth} \label{sec.figure of merit}
Having estimated errors on parameters in the fiducial Euclid configuration, in this section we move to investigating how well the growth of structure is measured when varying essential survey parameters, such as the observed area and the corresponding galaxy number densities observed, as well as the redshift range where galaxies are observed. To quantify the success in measuring growth, we resort to a dedicated Figure of Merit, analogous to the Dark Energy Task Force FoM.
In the Dark Energy Task Force report \citep{Albrecht:2006um}, a FoM  was introduced to quantify progress in measuring the properties of dark energy. This was defined as the inverse of the area enclosed in the $95\%$ confidence level contour of the equation of state parameters $w_0 - w_a$.

\subsection{Definition of the FoM}
Here we introduce an analogous FoM aimed at quantifying the precision of the growth measurement. Our FoM is defined as the inverse of the area enclosed in the $95\%$ c.l. contour of the growth parameter ($\gamma$ or $\mu_s$) and $\Omega_{m}$. We choose $\Omega_{m}$ as our second  parameter because we want the FoM to be usable for all models,  while e.g. the equation of state, which could represent an alternative to $\Omega_{m},$ is fixed to $-0.95$ in qLCDM and therefore cannot be used. Moreover, our main observable, the RSD, constrains $\Omega_{m}$ very accurately.
The FoM as we define it can be computed through this simple formula:
\be \label{eq.FoM_def}
{\rm FoM} =  \frac{\sqrt{\det F_{\gamma\Omega_{m}}}}{-2\pi \,\ln \left[{\rm erfc} \left(\frac{2}{\sqrt{2}}\right)\right]}\,.
\ee
or through an analogous one for $\mu_s$.

\subsection{Dependence of the FoM on survey specifications}
We compare here $\gamma$ and $\mu_s$ and explore three survey parameters: the survey area, corresponding to an associated galaxy number density, the maximum redshift $z_{max}$ and the redshifts range $z_{min}-z_{max}$.

If we wanted to vary the survey area independently of the other survey parameters, we would find a simple proportionality relation:  FoM $\propto$ area, independently of the parameterisation assumed. This can be easily understood by looking at Eq.~(\ref{eq.fisher_Tegmark}) and knowing that  $V_{surv}$ is always proportional to the total survey area, which means that $F_{ij} \propto $ area. Therefore, $\det F_{\gamma\,\Omega_{m}} \propto {\rm area}^2$ and from Eq.~(\ref{eq.FoM_def}), FoM $\propto$ area.

A more realistic question to be asked when planning a survey is rather whether it is more convenient to invest the total survey duration by mapping a smaller area for a longer time, so to increase the number density of observed galaxies, or rather mapping a larger region of the sky but collecting a sparser sample of galaxy spectra.

To understand this we have built a simple linear scaling on how the number of galaxies per redshift bin increases when the area is reduced and the exposure time per sky patch is increased but the total available time is fixed.
In particular, the same end-to-end simulations described in Sec. \ref{sec.simulations} provide us also with an estimate of the number density increase obtained when increasing the observing time per observed patch by $25\%$.
We assume that this increase corresponds to a $25\%$ decrease in survey area covered, given that the total survey time available is constant, and that this relation is linear.
From these considerations, we derive the following formula:
\be \label{eq.trade-off}
n(z) = n_{15000}(z)\left[ (1-C(z)) + C(z) \frac{15,000 \,{\rm deg}^2}{area} \right]\,,
\ee
where $n_{15000}$ is the number density corresponding to an area of $15,000$ square degrees and the values of $C(z)$ are listed in Table \ref{tab.trade-off}.
More precise answers would be obtained using e.g. the method of \citet{Bassett:2004np} but we expect this approximation to be sufficient at this stage to roughly describe the trade-off between area and number density.
\begin{table}\caption{Trade-off between number density and area: $C(z)$ for each redshift bin, to be inserted in Eq. (\ref{eq.trade-off}), in order to relate the increase in number density to variations in the total survey area.}
\centering
\begin{tabular}{c@{\hspace{0.4cm}} c@{\hspace{0.4cm}}  }
\\
\hline\hline\\
$z$ 	&  $C(z)$ \\
\hline
$0.7$	& $0.365786$\\
$0.8$	& $0.400569$\\
$0.9$	& $0.408442$\\
$1.0$	& $0.412037$\\
$1.1$	& $0.418330$\\
$1.2$	& $0.418001$\\
$1.3$	& $0.422914$\\
$1.4$	& $0.422946$\\
$1.5$	& $0.439024$\\
$1.6$	& $0.490589$\\
$1.7$	& $0.506262$\\
$1.8$	& $0.521747$\\
$1.9$	& $0.531328$\\
$2.0$	& $0.551155$\\
\hline\hline
\end{tabular}\label{tab.trade-off}
\end{table}
\begin{figure*}
\begin{center}
\includegraphics[width=0.45\textwidth]{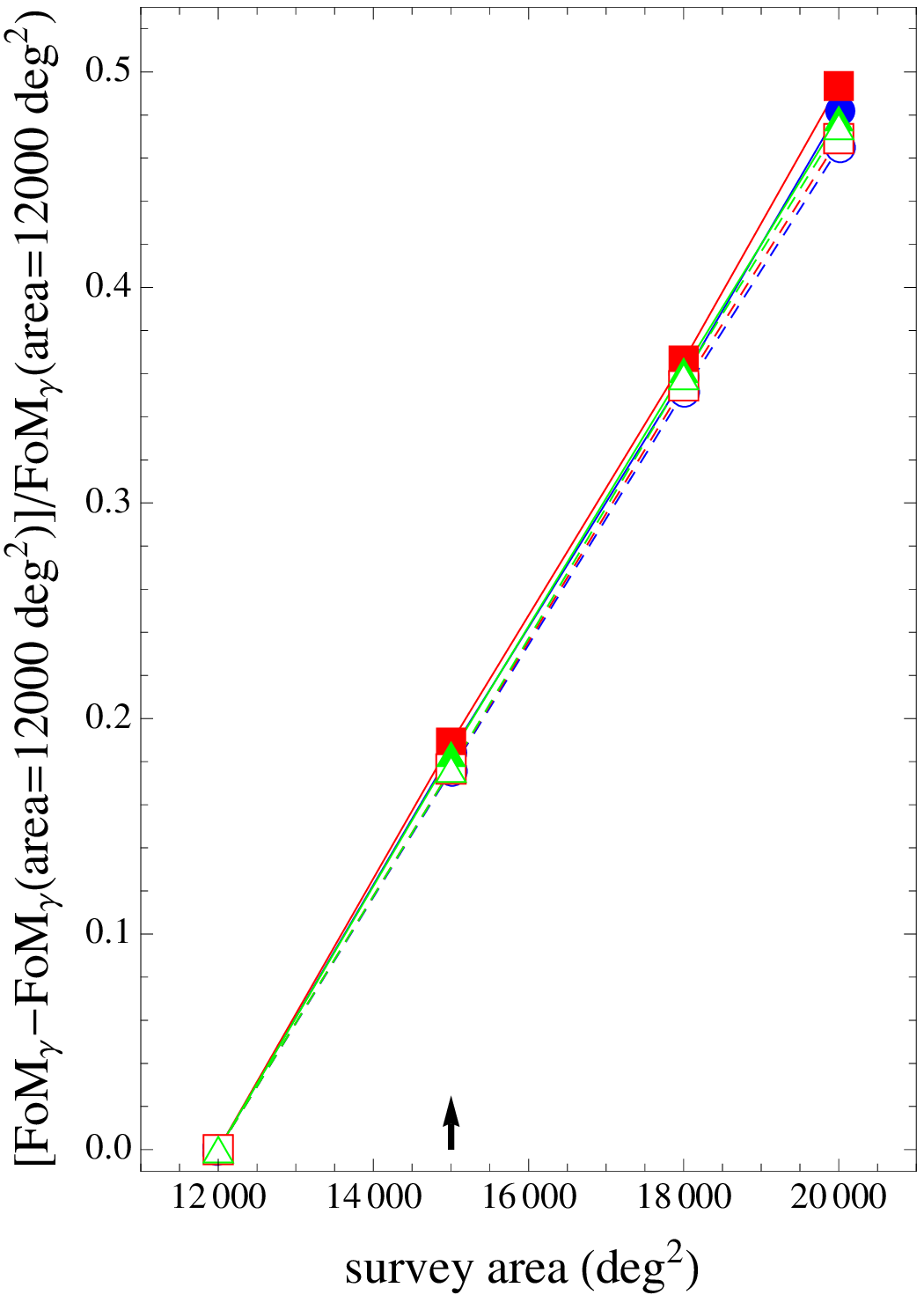}\quad
\includegraphics[width=0.45\textwidth]{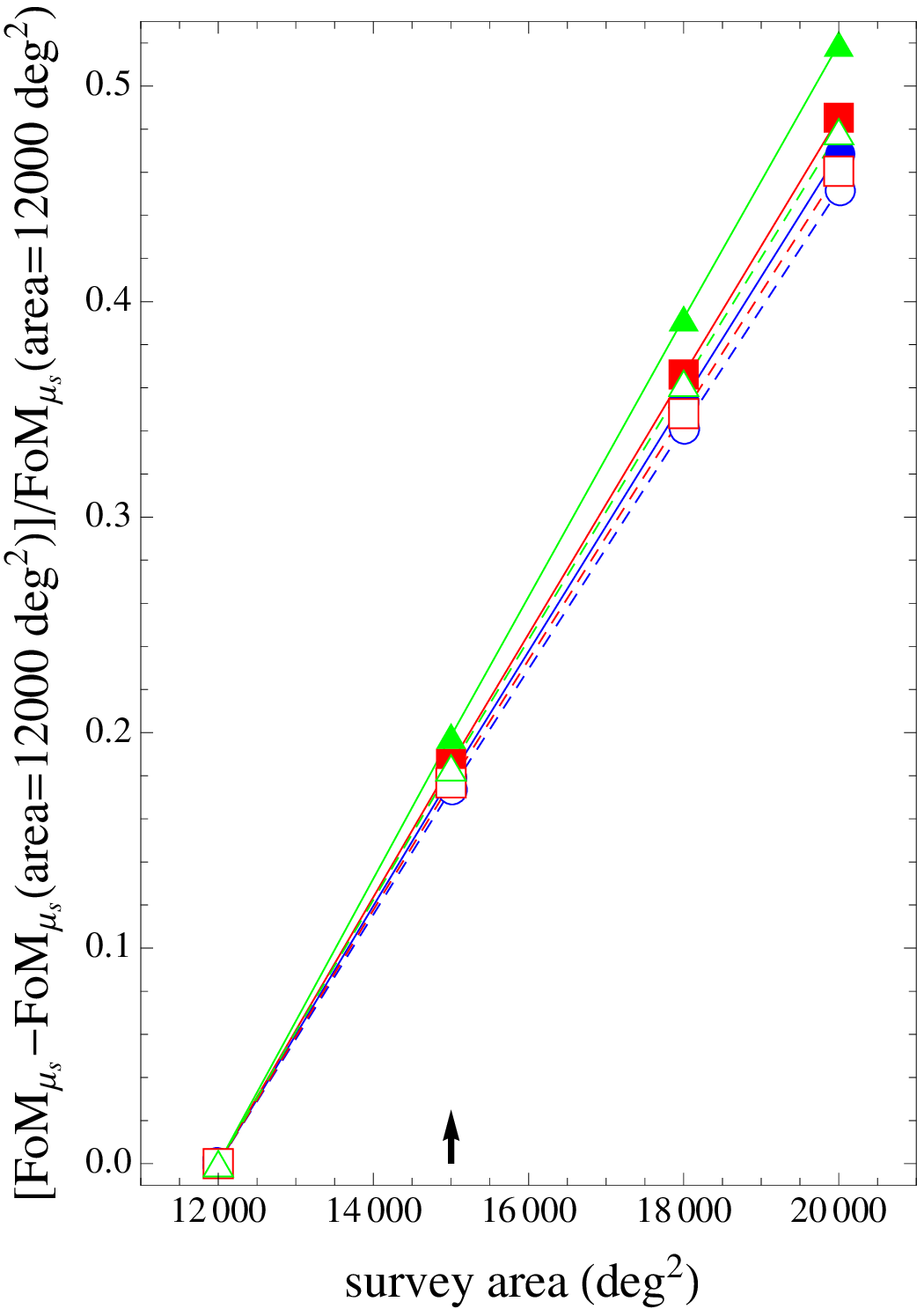}
\caption{Dependence of the relative FoM on the survey area and corresponding galaxy number density. The FoM is computed for the growth parameters $\gamma$ (left panel) and $\mu_s$ (right plot) and normalised to the FoM for ${\rm area} = 12000 {\rm deg}^2$: $ \left({\rm FoM} - {\rm FoM}({\rm area} = 12000 {\rm deg}^2)\right)/{\rm FoM}({\rm area} = 12000 {\rm deg}^2)$. Blue circles, red squares and green triangles are for qLCDM, wCDM and CPL respectively. Full (empty) symbols represent flat (curved) models. Solid (dashed) lines join flat (curved) models. The arrow indicates the baseline area of the Euclid survey, $15,000\,{\rm deg}^2$, resulting from an optimisation of the joint spectroscopic and photometric surveys.}
\label{fig.fom_area-density}
\end{center}
\end{figure*}
In Figure~\ref{fig.fom_area-density} we show the dependence of the FoMs for $\gamma$ and $\mu_s$ (left and right panels respectively) on the area - galaxy number density. In order to better appreciate the functional behaviour, we plot the relative improvement in the FoM when varying the area with respect to the FoM of area $ = 12,000 {\rm deg}^2$: $ \left( {\rm FoM(area)}-{\rm FoM} ({\rm area}=12000 {\rm deg}^2 )\right)/ {\rm FoM} ({\rm area}=12000 {\rm deg}^2 )$. 
Here we can see that the improvement in the growth measurement is nearly linear, at least for the area interval considered ($12,000 {\rm deg}^2 -20,000 {\rm deg}^2$), and mildly dependent on the dark energy model considered, but also rather slow, so that a large increase in area might not be favoured in case the cost for obtaining it grows too fast.  
Comparing left to right plot, we also note that in the case of the $\mu_s$ parameterisation the improvement of the FoM depends more strongly on the cosmological model one has previously assumed.  The largest improvement appears for flat CPL.

\begin{figure*}
\begin{center}
\includegraphics[width=0.45\textwidth]{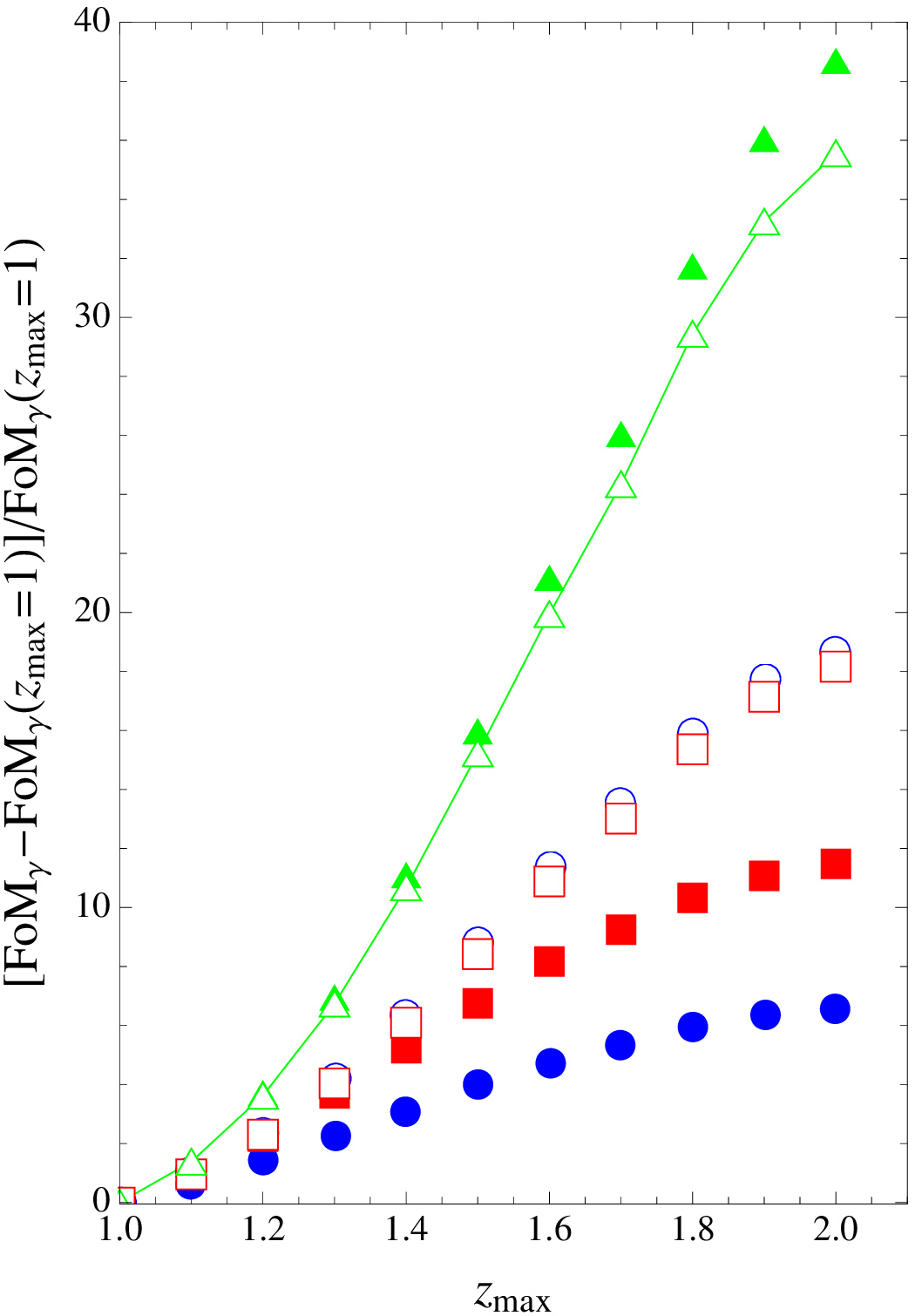}\quad
\includegraphics[width=0.45\textwidth]{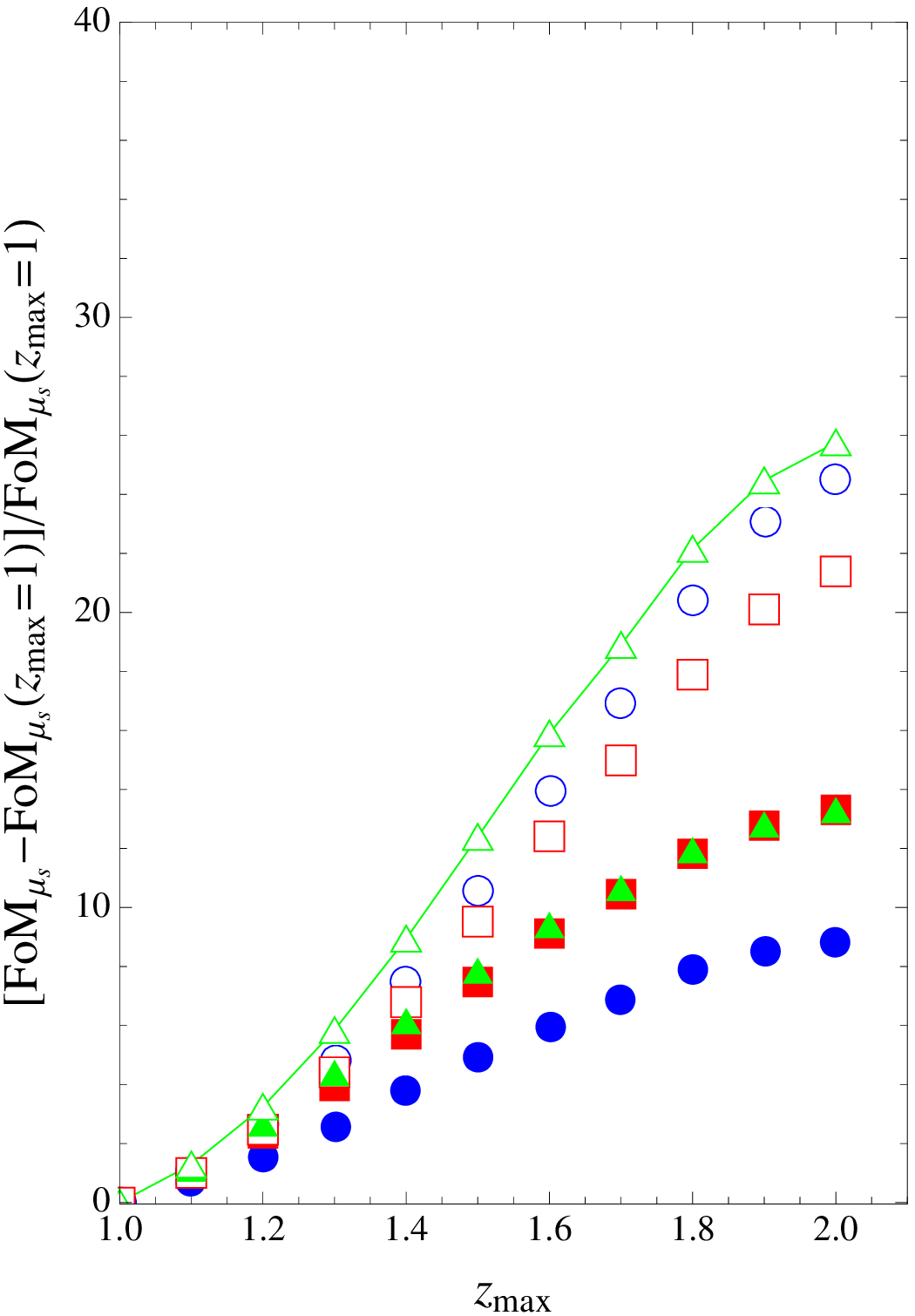}
\caption{Dependence of the FoM on the upper redshift limit of the survey $z_{max}$ (fixing the lower redshift limit $z_{min}$ to its minimum $0.7$), computed for the growth parameters $\gamma$ (left panel) and $\mu_s$ (right plot) and normalised to the FoM at $z_{max} = 1.0$. The symbols have the same meaning as in Fig.~\ref{fig.fom_area-density} but here the green solid line joins symbols corresponding to the most general model, curved CPL.}
\label{fig.fom_ratio_zmax_gamma}
\end{center}
\end{figure*}
 As regards the dependence of the FoM on $z_{max}$, we evaluated it by calculating our total Fisher matrix as the sum of the Fisher matrices computed for the redshift bins from $z = 0.7$ to $z = z_{max}$ only (see Sec. \ref{sec.Fisher_matrix_formalism} for further details on the general Fisher matrix calculation technique). Fig. \ref{fig.fom_ratio_zmax_gamma} shows the relative difference of the FoM($z_{max}$) to the FoM for $z_{max} = 1$: $ \left({\rm FoM}(z_{max}) - {\rm FoM}(z_{max} = 1)\right)/{\rm FoM}(z_{max} = 1)$. Here the amplitude of the spread due to model assumptions is much  wider than for the area - galaxy number density dependence, and wider in the case of $\gamma$. The latter is because the derivative of our observable $f \sigma_8$ with respect to $\mu_s$ decreases with redshift while its derivative with respect to $\gamma$ increases (as we have checked numerically), so that at $z \simeq 1$ $f \sigma_8$ is more sensitive to changes in $\gamma$ than in $\mu_s$. This explains why in the case of $\gamma$ and for the CPL model, which has the strongest time-dependence of the background cosmology, there is more advantage from higher redshift bins than for the case of $\mu_s$. 
In both parameterisations complex models gain more from higher redshift data, so e.g. by increasing $z_{max}$ wCDM's FoM improves more  than qLCDM's does, and the FoM of models with $\Omega_k \neq 0$  improves more than the corresponding one for flat models (with the exception of CPL with $\gamma$). In the case of $\gamma$, this is easy to explain. We know that $f = \Omega_m(z)^\gamma$. In the case of a flat qLCDM model $\Omega_m(z)$ approaches $1$ as soon as matter starts dominating, and this happens already for small values of $z$. From such $z$ on, the value of $f$ becomes practically independent of $\gamma$  so that higher redshift data do not help constraining it anymore. If instead the model is more complex, e.g. $\Omega_k$ is allowed to vary away from $0$, then $\Omega_m(z) \simeq 1$ at larger $z$ so that increasing $z_{max}$ improves more the FoM. Something similar also likely happens for $\mu$, but it is more difficult to illustrate it since we only have numerical solutions for $f$ in this case. 
Note also that for $\mu_s$ there is more difference between curved and flat models than for $\gamma$.

This and the fact that the FoM grows rapidly when increasing $z_{max}$ encourages the effort to reach the highest possible maximum limiting redshift in future surveys.

\begin{figure*}
\begin{center}
\includegraphics[width=0.45\textwidth]{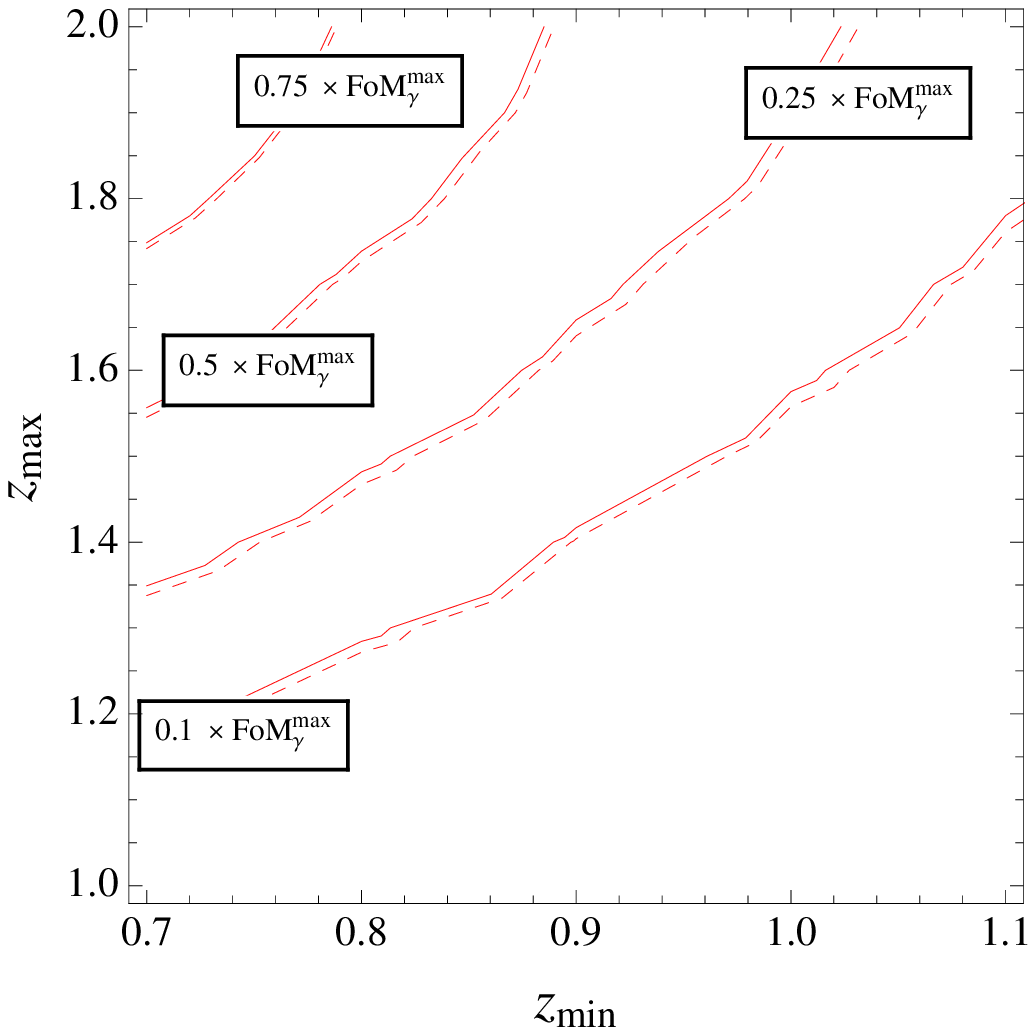}\quad
\includegraphics[width=0.45\textwidth]{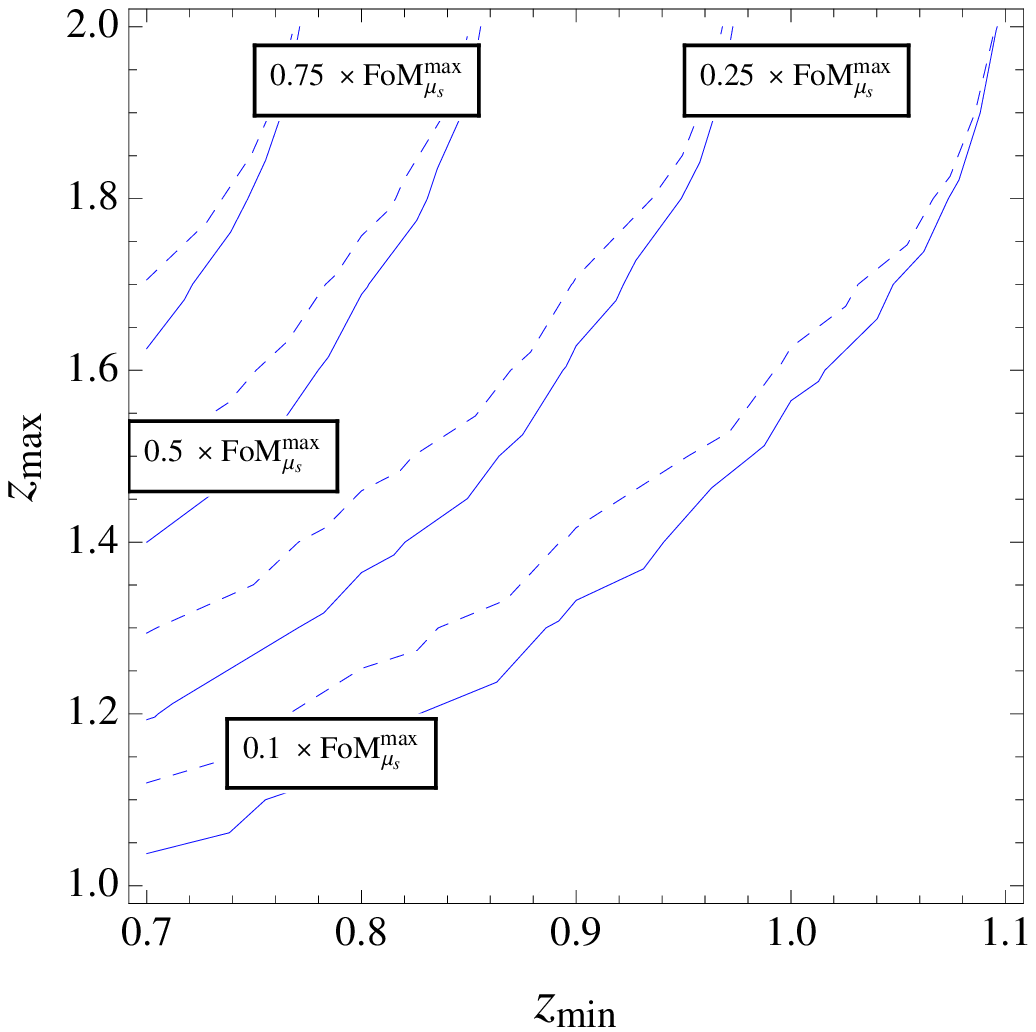}\quad
\caption{Dependence of the growth FoM  on the redshift range for the two different growth parameterisations considered: contours of FoM $= 10\%$, $25\%$, $50\%$ and $75\%$ of the maximum FoM reached in each case (corresponding to the use of the maximum redshift range $0.7 \leq z \leq 2.0$) are shown as a function of the lowest ($z_{min}$) and highest ($z_{max}$) redshift bins used. Here, a CPL model (dark energy with variable equation of state $w= w_0 + w_a(1-a)$) is assumed. Solid (dashed) lines correspond to a cosmology with $\Omega_k = 0$ ($\Omega_k \neq 0$) {\bf Left panel:} $\gamma$  parameterisation. {\bf Right panel:} $\mu_s$ parameterisation.}
\label{fig.fom_zmin-zmax}
\end{center}
\end{figure*}

Finally, we test the dependence of the growth FoM on the redshift interval covered by the survey. We plot in Figure \ref{fig.fom_zmin-zmax} contours of constant FoM (in particular, ${\rm FoM} = 10\%$, $25\%$, $50\%$ and $75\%$ of the maximum FoM reached in each case) as functions of  $z_{min}$ and $z_{max}$, where $z_{min}$ varies between $0.5$ and $1.1$ and $z_{max}$ between $0.9$ and $2.0$.
We concentrate here only on CPL dark energy, since this is the most complex and complete of our set of models. We first note that for the $\mu_s$ parameterisation there is stronger dependence on curvature of the FoM than for the $\gamma$ parameterisation (compare solid and dashed lines).
We also note that the contours for $\gamma$ and $\mu_s$ have different shapes. Let us consider a specific example.
If we look at $z_{min} \sim 0.75$, we need a $z_{max} \gtrsim  1.6$ to reach $50\%$ of the maximum FoM, for both parameterisations.  For other values of $z_{min}$, the two parameterisations give different indications. For $z_{min} < 0.75 $ ($> 0.75$), the minimum value of $z_{max}$ required in order to obtain $50\%$ of the maximum achievable FoM is larger (smaller) for $\gamma$ than for  $\mu_s$. This study is warning us against optimising an experiment on the base of one parameterisation only. In the absence of a clear preference for one parameterisation over the other, an operative way to proceed in this particular case would be to choose the most conservative $z_{min}$ and $z_{max}$ limits, for which the desired FoM is achieved in both parameterisations.

To end this section on figures of merit, we list in Table~\ref{tab.best_FoM_gamma} and \ref{tab.best_FoM_ms}  the FoMs, reached using the maximum available redshift range $0.7 \leq z \leq 2.0$ for all combinations of cosmological models, both for the $\gamma$ and the $\mu_s$ parameterisation. We also look at how the FoM is improved when adding data from a low-$z$ galaxy survey (as may be BOSS) and further add Planck data. In square brackets we show how the FoMs are degraded in the pessimistic case of the galaxy number density being half that forecasted in \citet{Geach:2008us}.
\begin{table*}\caption{Figures of Merit for $\Omega_m$ and $\gamma$. Both the case of fixing $\Omega_k = 0$ (flat space) and allowing it to vary (curved space) were listed. All figures in square brackets represent the case of the galaxy number density being halved. The addition of other surveys at lower redshift was considered for all models, while the effect of adding Planck was computed only for one representative case, i.e. that of the most complex model (curved CPL).}
\centering
\begin{tabular}{l@{\hspace{0.4cm}} l@{\hspace{0.4cm}} l@{\hspace{0.4cm}} l@{\hspace{0.4cm}} l@{\hspace{0.4cm}} l@{\hspace{0.4cm}} }
\\
\hline\hline\\
$\gamma$	& {\bfseries Euclid}		& 			& {\bfseries + low-z data}		& 				& {\bfseries + Planck} 	\\
			& flat space			& curved space& flat space				&curved space		& curved space			\\
\hline\\
qLCDM 		& 	545 [361]		 	& 209 [135]	& 	561 [376]				& 221 [137]		&			\\
wCDM		&	217 [146]			& 74 [48]		& 	225 [153]				& 75 [49]			&			\\
CPL 			& 	35 [23]			& 30 [20]	 	&	35 [23]				& 30 [20]			& 141 [140]	\\
\hline\hline
\end{tabular}\label{tab.best_FoM_gamma}
\end{table*}

\begin{table*}\caption{Figures of Merit for $\Omega_m$ and $\mu_s$. Both the case of fixing $\Omega_k = 0$ (flat space) and allowing it to vary (curved space) were listed. All figures in square brackets represent the case of the galaxy number density being halved. The addition of other surveys at lower redshift was considered for all models, while the effect of adding Planck was computed only for one representative case, i.e. that of the most complex model (curved CPL).}
\centering
\begin{tabular}{l@{\hspace{0.4cm}} l@{\hspace{0.4cm}} l@{\hspace{0.4cm}} l@{\hspace{0.4cm}} l@{\hspace{0.4cm}} l@{\hspace{0.4cm}} }
\\
\hline\hline\\
$\mu_s$	& {\bfseries Euclid}		& 			& {\bfseries + low-z data}		& 				& {\bfseries + Planck} 	\\
		& flat space			& curved space& flat space				&curved space		& curved space			\\
\hline\\
qLCDM 	& 	244 [159]				& 93 [59]	& 		251 [165]		& 94 [60]			&\\
wCDM	&	82 [55]				& 28 [18]	& 		85 [58]		& 29 [18]			&\\
CPL 		& 	18 [13]				& 9 [6] 	&		19 [13] 		& 9 [6]			& 82 [82]\\
\hline\hline
\end{tabular}\label{tab.best_FoM_ms}
\end{table*}

\section{Distinguishing General Relativity from modified gravity models} \label{sec.bayesian evidence}
We finally turn to trying and answering the question ``is the Euclid spectroscopic survey able to distinguish between GR and modified gravity?'' To do this we use model selection tools (aimed precisely at  telling how strongly a set of data prefers a model over other models) from Bayesian statistics. 
In particular, we apply the method of \citet{Heavens:2007ka}, where the Fisher matrix approach is generalised to the context of model selection.  The Bayesian tool for model selection is the Bayes' factor $B$, defined as the ratio of probabilities of model $M'$ to model $M$, given the same data $D$, independently of the values assumed by the model parameters $\theta'$ or $\theta$:
\be
B = \frac{p(M'| D)}{p(M| D)} = \frac{p(M')}{p(M)} \frac{ \int d\theta' p(D|\theta', M') p(\theta'|M')}{ \int d\theta p(D|\theta, M) p(\theta | M)}\,
\ee
where $p(M| D)$ is the probability of model $M$ given the data $D$, $p(M)$ is the prior probability of model $M$ (i.e. the probability that model $M$ is true {\itshape before} the experiment is done), which is unknown, and we will assume $p(M)=p(M')$; $p(D|\theta, M)$ is what is usually called the likelihood function (i.e. the probability that the data are true given the model $M$ with parameters $\theta$); $p(\theta | M)$ is the prior probability of the model parameters $\theta$ (i.e. the probability distribution that one believes the model parameters have before the experiment is done).
As in \citet{Heavens:2007ka}, $M'$ is here a dark energy model well described by the CPL parameterisation, while $M$ is a modified gravity model whose background expansion can still be described by the same $H(z)$ of the CPL parameterisation, (not necessarily by the same fiducial parameters) but now the growth index $\gamma$ is fixed to a fiducial value different from the GR value of $0.545$. Our fiducial $M'$ cosmology is that of Eqs. (\ref{eq.fid.pars_in}). For our case the forecasted Bayes' factor is \citep{Heavens:2007ka}
\be
\langle B \rangle = (2\pi)^{-1/2} \frac{ \sqrt{\det F}}{ \sqrt{\det F'}} \exp \left(-\frac{1}{2} \delta \theta_\alpha F_{\alpha \beta} \delta\theta_\beta \right) \Delta \gamma
\ee
where $\Delta\gamma$ corresponds to a uniform prior on $\gamma$, $F'$ and $F$ are the Fisher matrices for GR and the modified gravity model, respectively, and $\delta\theta_\alpha$ represent the shifts of the parameters of $M'$ due to a shift in $\gamma$ from the fiducial value $0.545$:
\be
\delta \theta_\alpha = - \left( F'^{-1} \right)_{\alpha \beta} G_{\beta} \left(\gamma-\gamma_{GR}\right) 
\ee
 for $\alpha$ corresponding to all parameters but $\gamma$, while 
$\delta \theta_\alpha = \left(\gamma-\gamma_{GR}\right) $
for $\alpha$ associated to the $\gamma$ parameter.
Here $G_\beta$ is the vector drawn from $F$ by extracting the column corresponding to the parameter $\gamma$ and the rows corresponding to all parameters except $\gamma$.
From the above we see that $\langle B \rangle$ depends on the offset of $\gamma$  with respect to the GR value, $(\gamma-\gamma_{GR})$, and on the prior on $\gamma$, $\Delta \gamma$. 

\begin{figure*}
\begin{center}
\includegraphics[width=0.7\textwidth]{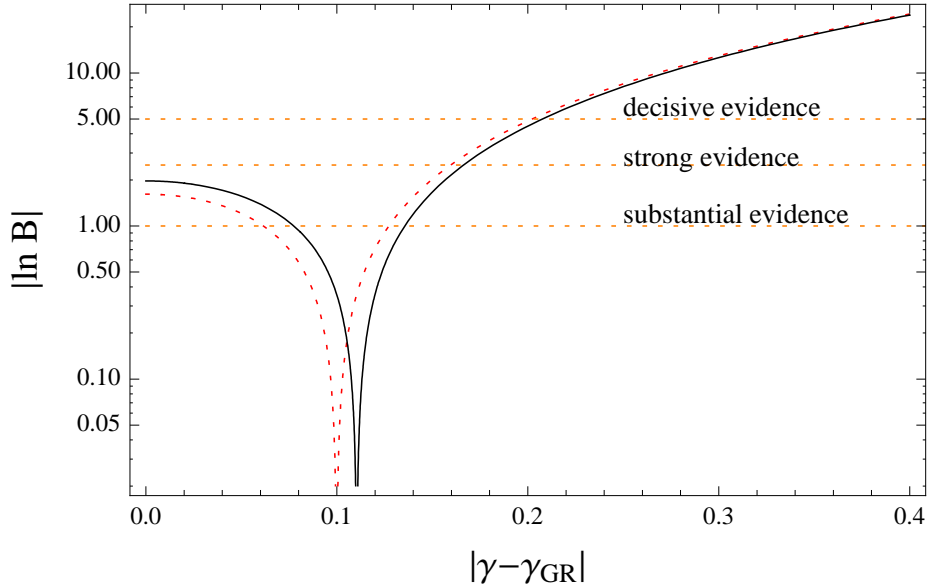}
\caption{Dependence of 
$\ln \langle B \rangle$ on $|\gamma-\gamma_{GR}|$ for different priors in the Euclid spectroscopic galaxy survey. The solid black line corresponds to a  uniform prior distribution for $\gamma$ with $\Delta \gamma = 1.0$, the red dotted line to $\Delta\gamma = 0.7$. The dotted red lines correspond to $\ln B=1 $, $\ln B =2.5 $ and $\ln B=5$, delimiting the regions where evidence in favour of one model with respect to the other is 'substantial', 'strong' and 'decisive' according to Jeffreys' scale. 
The cusp corresponds to the case where $B = 0$, i.e. there is no evidence in favour of one model with respect to the other. This means that to the left of the cusp GR is favoured with respect to modified gravity models, while to its right modified gravity models are favoured.
}
\label{fig.evidence}
\end{center}
\end{figure*}

In Figure \ref{fig.evidence} we show the dependence of the log of  $\langle B \rangle$ on $|\gamma -\gamma_{\rm GR}|$ for two different priors (represented by the solid and red dashed lines) in the case of our Euclid-like spectroscopic galaxy survey. The horizontal dotted lines describe values of $\ln B$ which correspond to  'substantial' ($1<\ln B<2.5 $), 'strong'  ($2.5<\ln B<5 $) and 'decisive'  ($\ln B>5 $)  evidence in favour of one model with respect to the other (bottom to top) according to Jeffreys' scale \citep{jeff}. As can be seen from the figure, the Euclid spectroscopic survey alone will be able to substantially distinguish between GR and a modified gravity model if $\gamma -\gamma_{\rm GR} \geq 0.13$, while it will be able to decisively distinguish between models if $\gamma -\gamma_{\rm GR}\geq 0.2$.
 We have also computed the evidence using the $\mu$ parameterisation, with uniform prior distributions in the interval $\Delta \mu_s = 3$ and $\Delta \mu_s = 5$. For both priors it results that with Euclid spectroscopic data alone 'substantial'  ('strong') evidence can be obtained in favour of a modified gravity if the latter has $\mu_s \gtrsim 1$ ($\mu_s \gtrsim 1.4$).
The addition of the weak lensing data from the Euclid photometric survey is expected to improve these results considerably.

\section{Conclusions}\label{sec.conclusions}
In this paper we have investigated how strongly the Euclid galaxy spectroscopic survey in the current reference configuration can constrain the growth of structure and consequently how well it can differentiate a GR cosmology from alternatives to it.

We have found that we can reach precisions between $1.3\%$ and $4.4\%$ in the measurement of $f \,\sigma_8$ depending on the redshift bin, where the highest precision is reached for $z\simeq 1.0$. 

Comparing the Euclid spectroscopic survey with other ongoing and future galaxy redshift surveys we note that Euclid will reach the highest precision in the growth rate measurement. Euclid will be perfectly complementary to BOSS and BigBOSS: the three surveys together will allow to cover an extremely large redshift range: $0.1<z<3.5$.

This precision in $f\,\sigma_8$ translates into a precision in the measurement of the growth index $\gamma$ which depends on the specific background cosmology adopted. We have obtained marginalised errors on $\gamma-\Omega_m$ (or $\gamma-w_0$) between $5\%$ and $10\%$.  The parameterisation of the growth rate $f$ we have adopted is $f= \Omega_m^\gamma$ (for curved space, $f= \Omega_m^\gamma +(\gamma -4/7)\Omega_k$), where a departure from GR is represented by a deviation of $\gamma$ from $0.545$. We have considered nested background models: qLCDM (a model with constant $w = -0.95$), wCDM and CPL, both flat and curved. 

We have compared the relative gain in growth FoM (quantifying the precision in the joint measurement of $\Omega_m$ and $\gamma$) for two different growth parameterisations, being the already mentioned $\gamma$ and the parameter $\mu_s$ \citep{Pogosian:2010tj, Song:2010fg}. 
We have found that when increasing the survey area (and correspondingly reducing the galaxy number density, having fixed the total observing time) the FoM grows linearly. Moreover, this growth is quite mild. 

The relative gain in FoM when increasing $z_{max}$ is large, both using $\mu_s$ and  $\gamma$. We also have noted that (for curved models) the relative improvement increasing $z_{max}$ is approximately linear, encouraging the effort to reach the maximum possible limiting redshift. We then have examined the dependence of the FoM on the redshift interval covered by the survey. From Fig. \ref{fig.fom_zmin-zmax} it is possible to notice that to reach a desired FoM improvement one needs lower $z_{min}$ (or higher $z_{max}$) when considering $\mu_s$ than when choosing $\gamma$. This warns us against relying on one parameterisation only when optimising an experiment.

Finally we have forecasted the Bayesian evidence for Euclid and found that the spectroscopic survey alone will be able to substantially (decisively) distinguish between GR and modified gravity models having $\gamma-\gamma_{\rm GR} \geq 0.13$ ($\geq 0.2$). This result is expected to improve even further when adding data from the photometric survey of Euclid, which improve noticeably the precision in the measurement of $\gamma$ (see e.g. Fig.2.5 of \citealt{EditorialTeam:2011mu}).

\section*{Acknowledgements}
We wish to thank Domenico Sapone for essential and kind help in developing the Fisher code, Cinzia Di Porto for kindly providing details on coupled models,  Ben Granett, Matteo Chiesa, Davide Bianchi and Alida Marchetti for enlightening discussions. EM was supported by INAF through PRIN 2007, by the Spanish MICINNs Juan de la Cierva programme (JCI-2010-08112), by CICYT through the project FPA-2009 09017, by the Community of Madrid  through the project HEPHACOS (S2009/ESP-1473) under grant P-ESP-00346 and by the  European Union FP7  ITN INVISIBLES (Marie Curie Actions, PITN- GA-2011- 289442). WJP is grateful for support from the UK Science and Technology Facilities Council (STFC). LS is grateful for support from the
Georgian National Science Foundation grant GNSF ST08/4-442 and SNFS SCOPES grant no.
128040. LS and WJP are grateful for support from the European Research Council (ERC). YW was supported in part by DOE grant DE-FG02-04ER41305. CC acknowledges PRIN MIUR  ``Dark energy and cosmology with large galaxy surveys''. EM, LG, BG, PF, ER, AC, CC, NR and GZ acknowledge support from ASI Euclid-NIS I/039/10/0 grants.

\bibliographystyle{mn2e}
\bibliography{growth_FOM_v4.0}

\appendix
\section{Growth of structures}\label{sec.growth of structures}
In a Friedmann-Lemaitre-Robertson-Walker (FLRW) universe, the growth of
linear perturbations is described by the perturbed Einstein equations
and conservation of the energy-momentum tensor equations. If we consider
scalar perturbations about a FLRW background in the Newtonian gauge, in Fourier space and in the limit of small scales, an equation for the growth rate $f$ can be derived:
\be \label{eq.f_evol}
\frac{d f}{d\ln a} + f^2 + \left(  2 + \frac{d\ln H}{d\ln a} \right)f = \frac{3}{2}\Omega_m\,.
\ee
A solution for $f$ has now to be found and depends on the form of $H$.

\subsection{Parameterisations of growth in GR and deviations from it: a constant growth index}
\subsubsection{Flat LCDM and quintessence}
If we assume a flat model with a simple cosmological constant or a dark energy with (quasi) constant equation of state $w$,  with Friedmann equation $H^2 = H_0^2[ \Omega_{m}a^{-3} + (1-\Omega_{m})a^{-3(1+w)}]$, then it is possible to find a good solution of Eq.~(\ref{eq.f_evol}) by assuming
\be \label{eq.f_simple-param}
f = \Omega_m (a)^\gamma\,,
\ee
where the growth index $\gamma$  is a constant. This was first
proposed by \citet{peebles} for a matter-dominated universe at $z = 0$,
with $\gamma \simeq 0.6$. A better approximation for the
same model was found by \citet{fry} and \citet{lightman}, with $\gamma =
4/7$. 
For the case of LCDM, $\gamma = 0.545$ is a very good approximation, at least in the redshift range of interest here.
As shown in \citet{Wang:1998gt}, if we consider, instead of a standard cosmological constant, a quintessence field with slowly varying  $w$, then $\gamma = 0.545$ is still a good approximation, since the correction term has a weak dependence on $w$ if this is close enough to $-1$.

For larger redshift ranges, from $z_{CMB}$ to today, and demanding a very high precision, a different solution for LCDM was proposed by \citet{Ishak:2009qs}.
For models with non-slowly varying $w$,  \citet{Linder:2007hg} and \citet{Linder:2005in} propose different parameterisations.

We restrict ourselves to models with slowly varying $w$ and decide not to use the large redshift range parameterisation since we do not require an accuracy that high. Instead, we are interested in parameterisations for models where curvature is present, since this parameter has been shown to be degenerate with dark energy and we wish to explore it in a consistent way.

\subsubsection{Generalisation to curved LCDM and quintessence}
In the case of a curved LCDM model, the first attempts at modelling $f$ were
 proposed by \citet{martel} and \citet{lahav}.

More recently, in \citet{Gong:2009sp}, two possible approaches are suggested. The first (less accurate) approach consists in still assuming $f(z) = \Omega_m (z)^\gamma$, but changing the form of $\gamma$. The second consists in taking $f(z) = \Omega_\gamma (z)^\gamma + \alpha \Omega_k (z)$ (with $\Omega_k (z) = \rho_k (z)/\rho (z)$).

In the second case, which is found to be more accurate, the solution is
\be \label{eq.f_param_curv}
f(z) = \Omega_m^\gamma + (\gamma - 4/7) \Omega_k \,, \quad\quad\quad \gamma = 0.545 \,.\\
\ee
We use the parameterisation (\ref{eq.f_param_curv}) for our forecasts.

\subsubsection{Modified gravity: the case of DGP, $f(R)$ gravity and scalar-tensor theories} \label{subsec.mod-grav}
A parameterisation as that of Eq.~(\ref{eq.f_simple-param}) or (\ref{eq.f_param_curv}) is valid also for some modified gravity models  (see \citealt{Amendola:book} for a complete review of viable alternatives to GR), when taking a different fiducial $\gamma$. 

Let us first take the DGP model. For flat space, \citet{Linder:2007hg} showed that the growth rate $f$ is still well parameterised by Eq.~(\ref{eq.f_simple-param}) with a constant $\gamma$. \citet{Wei:2008ig} and \citet{Gong:2009sp} derived the expected value of $\gamma$ by solving the modified growth equations of \citet{Lu, Koyama:2005kd}: $ \gamma = 11/16$.
A constant $\gamma$ still fits the curve well enough \citep{Linder:2007hg}.
If we include curvature, we can proceed, following \citet{Gong:2009sp}, as for curved quintessence, e.g. parameterising $f(z)$ as in Eq.~(\ref{eq.f_param_curv}), and we find that the best fitting $\gamma$ is again  $\gamma = 11/16$.

Instead of adding extra dimensions, it is possible to modify gravity in four dimensions. This is done e.g. in $f(R)$ theories \citep{Capozziello:2002rd, Capozziello:2003gx, Carroll:2003wy}, where the Einstein-Hilbert Lagrangian density $R - 2 \Lambda$ is modified into a different function, $R + f(R)$. The function $f(R)$ is strongly constrained  by local gravity tests \citep{Amendola:2007nt}, nevertheless the phenomenology of the growth of structure for these models is still very rich. Their growth index $\gamma$ is in general time- and scale-dependent \citep{Tsujikawa:2007gd, Gannouji:2008wt, Tsujikawa:2009ku, Motohashi:2010qj}. As found in \citet{Tsujikawa:2009ku} and \citet{Gannouji:2008wt}, the importance of scale-dependent dispersion of $\gamma$ depends on the particular $f(R)$ model chosen and on its parameters. For some particular values of the parameters, there can be scale-independent growth. We do not consider scale dependence in this work, although it would be very interesting to examine this feature, as it might be a smoking gun for $f(R)$ models \citep{Pogosian:2010tj}. Time dependence can arise in general. Forecasts on $\gamma(a)$ for Euclid-like galaxy redshift surveys have been computed in \citet{DiPorto:2011jr} and we refer the reader to this paper for a full analysis of the topic. In this work we concentrate on a constant $\gamma$, which somehow corresponds to a $\gamma(a)$ averaged  over the redshift range of the survey, weighted by the error on $f$ in each redshift bin. Given that viable $f(R)$ models show a lower growth index $\gamma(z=0) \sim 0.4$ with respect to GR, becoming even smaller at higher $z$,  if one was to detect an unusually low averaged constant $\gamma$, this would point to $f(R)$ models. The use of a time and possibly scale dependent $\gamma$ could then in principle allow to distinguish among different $f(R)$ models.

Scalar-tensor theories  \citep{Amendola:1999qq, Uzan:1999ch, Chiba:1999wt, Bartolo:1999sq, Perrotta:1999am}  are a generalisation of $f(R)$ models, where the Lagrangian density is $1/2 \,f(\phi, R) - 1/2\,\zeta(\phi)(\bigtriangledown\phi ) $ .
Given the generality of these theories, the phenomenology of their growth of structure is very rich. The growth of matter perturbations in some of these models has been studied e.g. in \citet{DiPorto:2007ym, Gannouji:2008wt, Kobayashi:2010wa}. In general $\gamma$ depends on time (and scale), but there are also specific cases where $\gamma \sim {\rm const}$ for redshifts $\lesssim 2$ \citep{Kobayashi:2010wa}.

\subsection{Parameterisations of growth in GR and deviations from it: physical parameters} \label{app.growth-par}
Instead of parameterising the deviation from GR with $\gamma$, we can use a different approach, proposed by \citet{Amendola:2007rr}. This consists in directly parameterising the full Einstein equations instead of some already approximated version of them (e.g for small scales). This approach may turn useful when combining constraints from different observational tools, which might need different approximations. This way, the parameters would always have a physical meaning and one would avoid difficulties in interpreting results, as pointed out very neatly in \citet{Pogosian:2010tj}. It is precisely the parameterisation $(\mu, \,\eta)$ proposed in \citet{Pogosian:2010tj} that we use and compare to $\gamma$. The Einstein equations defining the parameterisation are:
\be
k^2 \Psi = - 4\pi G_N a^2 \mu (a, k) \rho\, \delta\,,\quad\quad\quad   \frac{\Phi}{\Psi} = \eta (a, k)\,,
\ee
where $\Psi$ and $\Phi$ are the metric perturbations in the Newtonian gauge:
\be
ds^2 = -a^2 \left[(1+2\Psi) d\tau^2 - (1-2\Phi) d\vec{x}^2\right]\,.
\ee
Since the galaxy power spectrum is sensitive to $\Psi$ and not to the anisotropic stress (which depends on the difference between the two potentials $\Psi$ and $\Phi$), the parameter $\mu$ is sufficient for our work and we do not use $\eta$ at all.

For simplicity, we decide to assume again scale-independence: $\mu = \mu(a) $ (although see \citealt{Pogosian:2010tj} and Sec.~\ref{subsec.mod-grav} for reasons to keep scale-dependence).
We model $\mu$ as in \citet{Song:2010fg}:
\be 
\mu(a) = 1+ \mu_s a\,,
\ee
which is motivated by DGP and reduces to GR for $\mu_s = 0$. So, using again the function $f$ defined previously, we obtain
\be 
\frac{d f}{d \ln a} + f^2 + \left( 2 + \frac{d \ln H}{d \ln a} \right) f - \frac{3}{2} \Omega_m  \left(1+\mu_s a \right) = 0\,.
\ee
We can solve this equation numerically, by imposing the initial condition at an initial redshift of matter domination $z_{md}$: $f(z_{md}) = \Omega(z_{md})^{\gamma}$ (where $\gamma$ can be determined from Eq.~(32) of \citet{Pogosian:2010tj}).

\section{Predictions using the $\mu$ parameterisation} \label{app.mu_s-forecasts}
 Here we present forecasts on the marginalised errors on cosmological parameters when, instead of using $\gamma$, the alternative $\mu$ parameterisation is used (see Appendix \ref{app.growth-par}).
\begin{figure}
\begin{center}
\includegraphics[height=0.23\textheight]{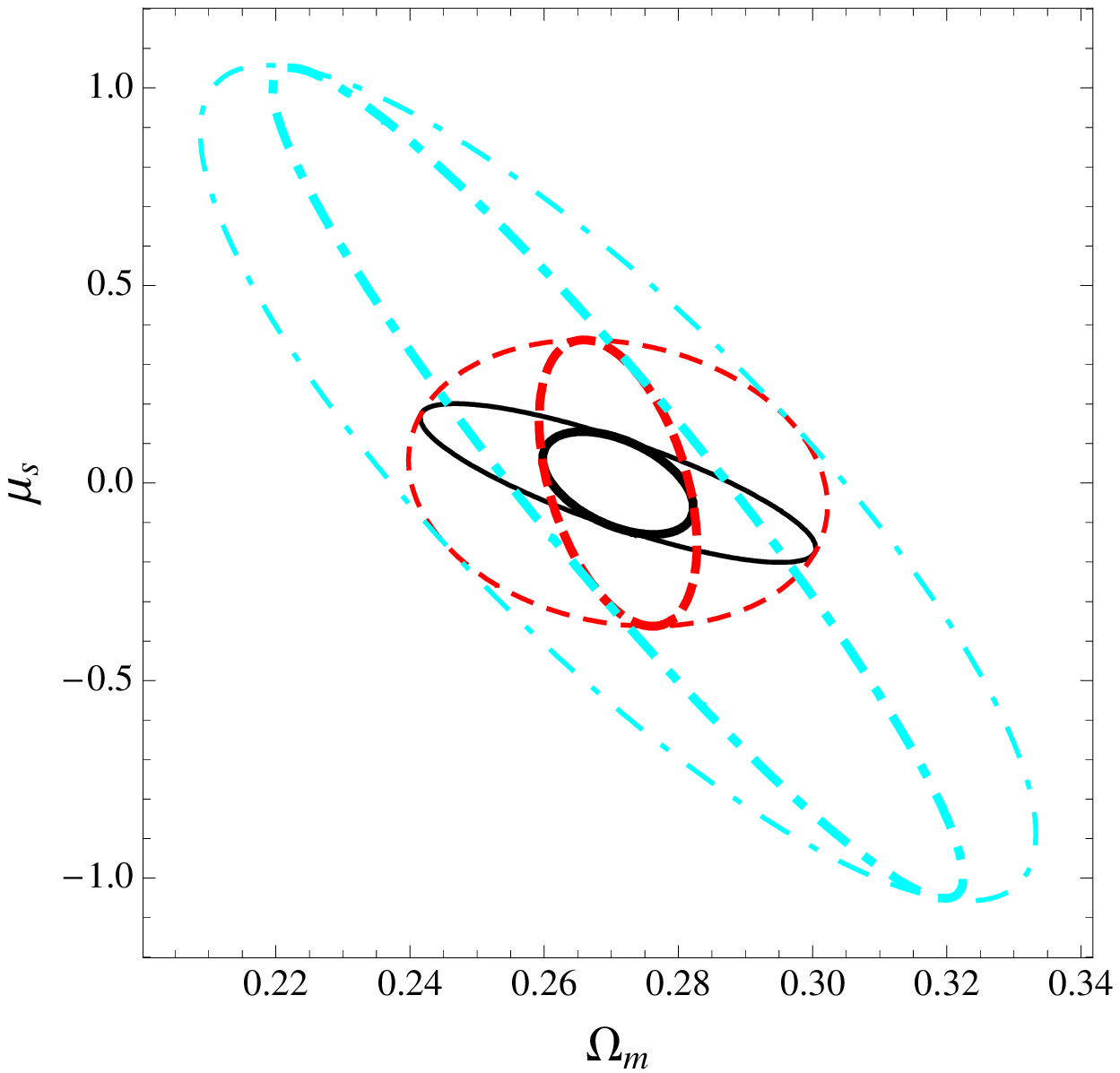}
\caption{Forecasted marginalised errors on $\Omega_m$ and $\mu_s$ using the Euclid spectroscopic survey. Solid (black) lines correspond to qLCDM, dashed (red) lines to wCDM and dot-dashed (cyan) lines to CPL. Thick lines represent models with $\Omega_k = 0$, while thin lines indicate curved models.}
\label{fig.fisher-growth-Om-mu_s}
\end{center}
\end{figure}
\begin{figure}
\begin{center}
\includegraphics[height=0.23\textheight]{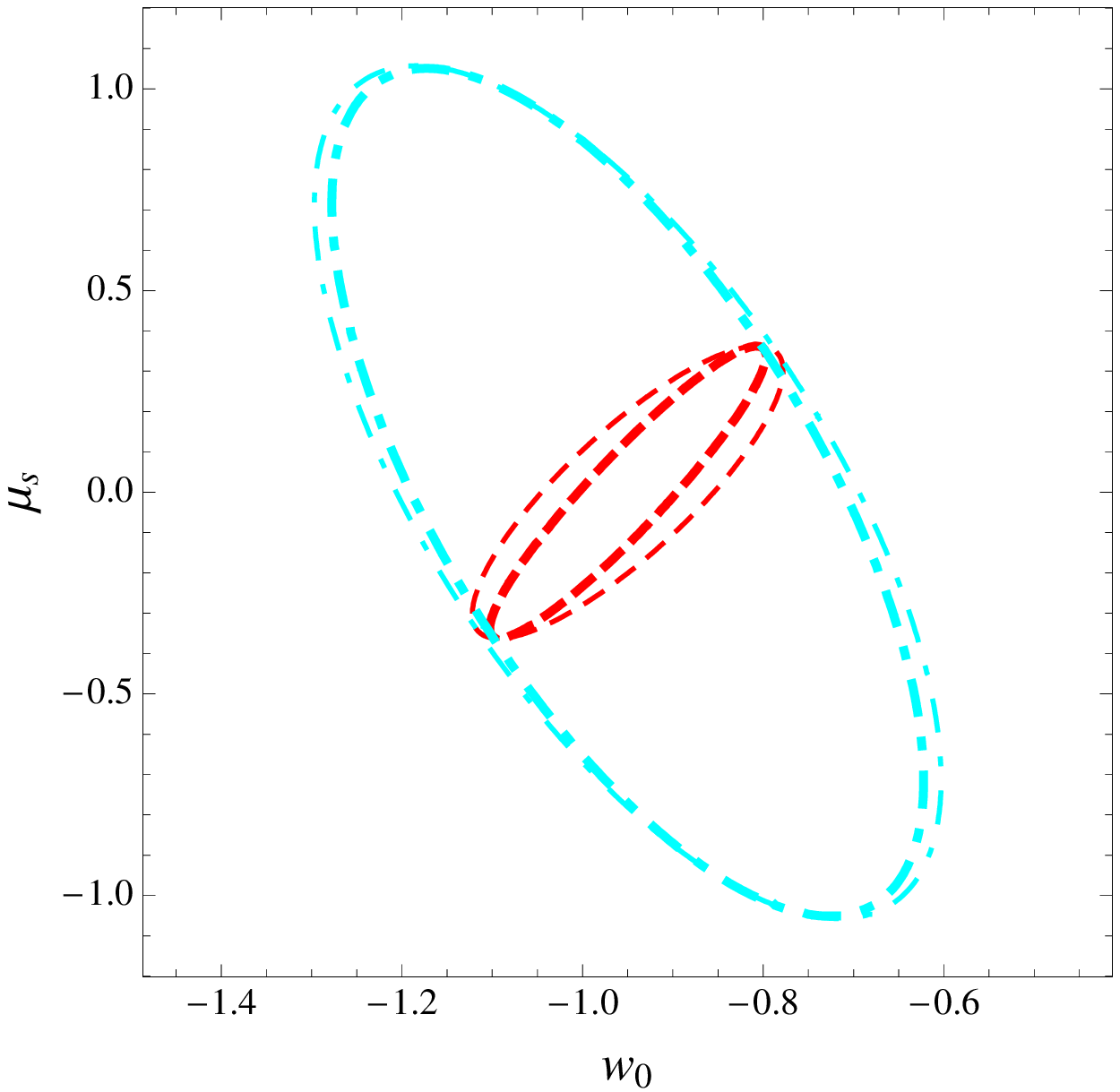} 
\caption{Forecasted marginalised errors on $w_0$ and $\mu_s$ using the Euclid spectroscopic survey.   Dashed (red) lines correspond to wCDM and dot-dashed (cyan) lines to CPL. Thick lines represent models with $\Omega_k = 0$, while thin lines represent curved models.}
\label{fig.fisher-growth-mu_s}
\end{center}
\end{figure}
As we can see from Figs. \ref{fig.fisher-growth-Om-mu_s} and \ref{fig.fisher-growth-mu_s}, with Euclid spectroscopic data alone the absolute marginalised error  on $\mu_s$ and $\Omega_m$ or $w_0$ will be $\sim 2$. \begin{figure}
\begin{center}
\includegraphics[height=0.23\textheight]{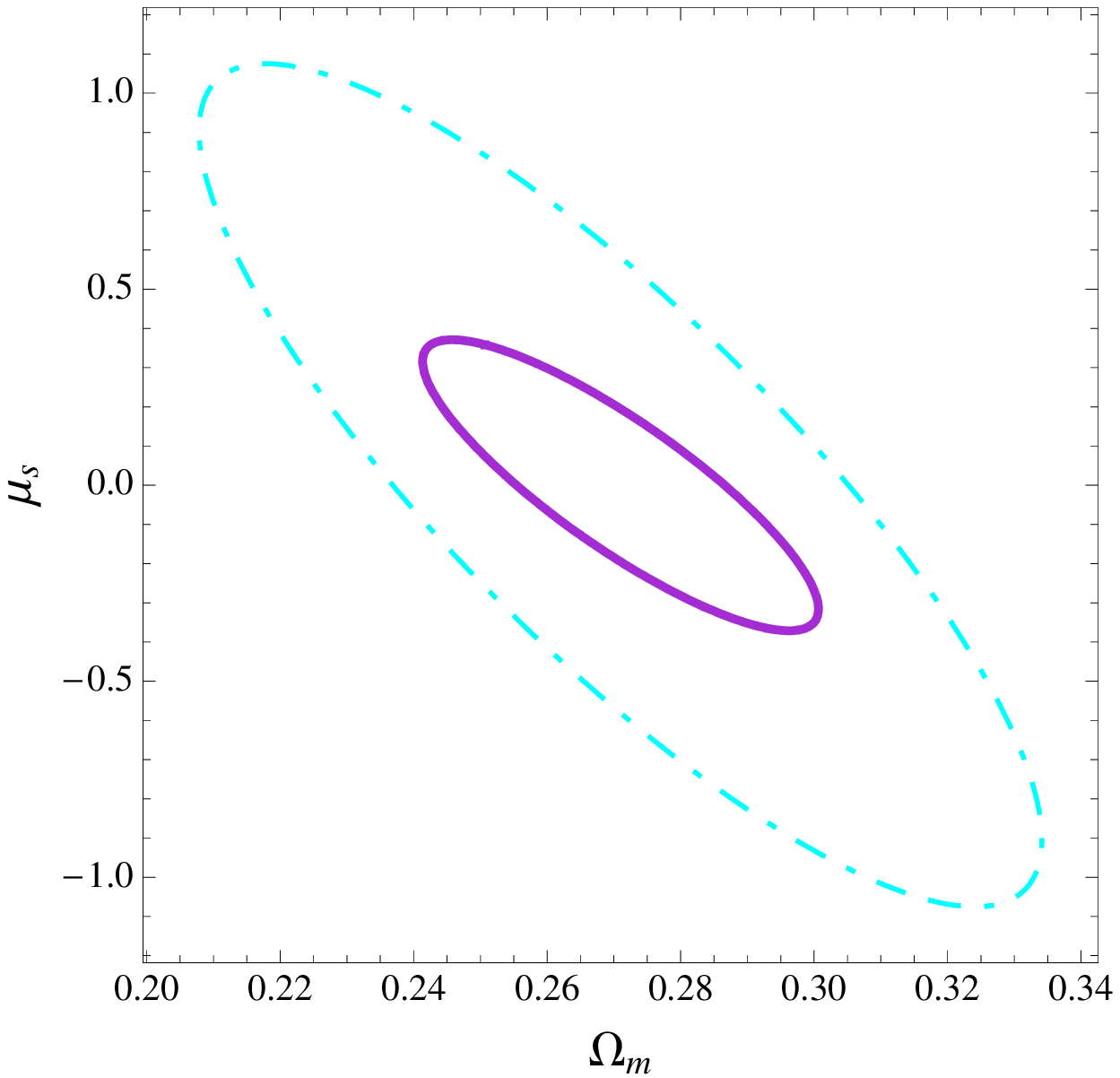}
\caption{Comparison of marginalised errors on $\Omega_m$ and $\mu_s$ in curved CPL, obtained using Euclid only and joint Euclid and Planck data. Dot-dashed (cyan) contours correspond to error ellipses using Euclid data only, while solid (purple) contours show joint constraints from Euclid and Planck.}
\label{fig.Planck_CPL_Om-ms}
\end{center}
\end{figure}
\begin{figure}
\begin{center}
\includegraphics[height=0.24\textheight]{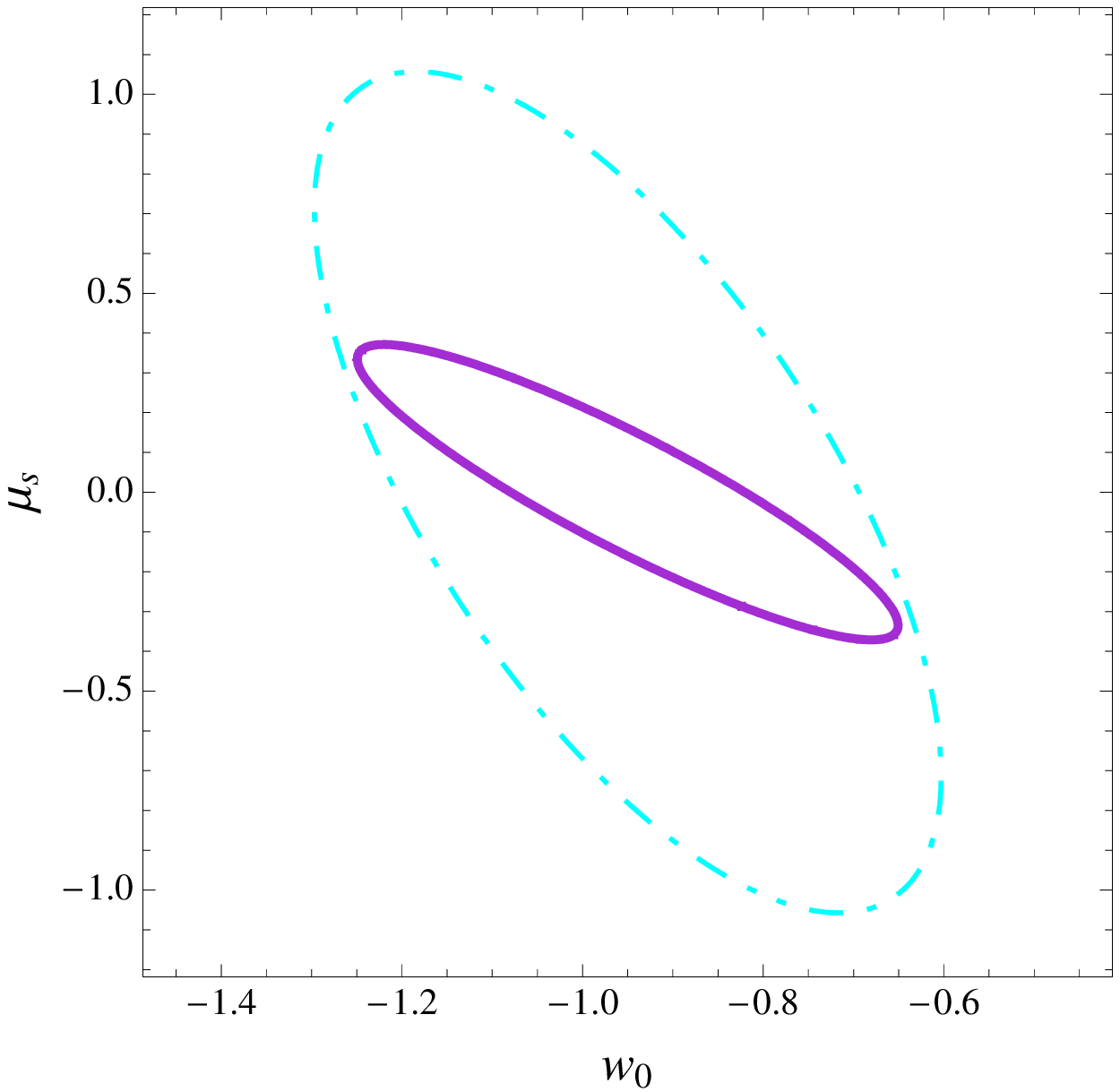}
\caption{Comparison of marginalised errors on $w_0$ and $\mu_s$ in curved CPL, obtained using Euclid only and joint Euclid and Planck data. Dot-dashed (cyan) contours correspond to error ellipses using Euclid data only, while solid (purple) contours show joint constraints from Euclid and Planck.}
\label{fig.Planck_CPL_ms}
\end{center}
\end{figure}
Adding Planck will improve the measurement of $\mu_s$ by more than $50\%$, reducing the error to $\sim 0.7$ (see Figs. \ref{fig.Planck_CPL_Om-ms} and \ref{fig.Planck_CPL_ms}).

\bsp

\label{lastpage}

\end{document}